\documentclass[reqno,a4paper]{amsart}
\usepackage{amsmath}
\usepackage{amssymb}
\usepackage{amsaddr}
\usepackage{geometry}
\usepackage{graphicx}
\usepackage[sort&compress,numbers]{natbib}

\usepackage{hyperref}
\hypersetup{
  unicode=false,                % nonLatin characters in Acrobat’s bookmarks
  pdftoolbar=true,              % show Acrobat’s toolbar?
  pdfmenubar=true,              % show Acrobat’s menu?
  pdffitwindow=false,           % window fit to page when opened
  pdfstartview={FitH},          % fits the width of the page to the window
  pdftitle={Compressing multireference character of wave functions via fermionic mode optimization},
  pdfauthor={Mih{\'a}ly M{\'a}t\'e} {Kl{\'a}ra Petrov} {{\"O}rs Legeza} {Szil{\'a}rd Szalay} ,
  pdfsubject={},
  pdfcreator={pdflatex},        % creator of the document
  pdfproducer={overleaf},            % producer of the document
  pdfkeywords={DMRG} {TNS} {MPS} {fermionic mode optimization} {molecular orbital optimization} {nitrogen dimer},
  pdfnewwindow=true,            % links in new window
  colorlinks=true,              % false: boxed links; true: colored links
  linktoc=page,                 % defines which part of an entry in the table of contents is made into a link
  linkcolor=blue,               % color of internal links       (red)
  citecolor=blue,               % color of links to bibliography
  filecolor=blue,               % color of file links
  urlcolor=blue                 % color of external links
}

\newcommand{\ket}[1]{\vert #1 \rangle}
\newcommand{\id}{\mathbb{I}}
\DeclareMathOperator{\Tr}{Tr}

\begin{document}

\title[Compressing multireference character via fermionic mode optimization]{Compressing multireference character of wave functions via fermionic mode optimization}

\author{Mih{\'a}ly M{\'a}t{\'e}$^{1,2}$,
Kl{\'a}ra Petrov$^{1}$,
Szil{\'a}rd Szalay$^{1,3,4}$,
{\"O}rs Legeza$^{1,5}$}

\address{$^1$Theoretical Solid State Physics Department,
Wigner Research Centre for Physics,
Konkoly-Thege Mikl{\'o}s str., Budapest, H-1121, Hungary}
\address{$^2$Department of Physics of Complex Systems,
E\"otv\"os Lor\'and University,
P.O. Box 32, Budapest, H-1518, Hungary}
\address{$^3$Department of Theoretical Physics,
University of the Basque Country UPV/EHU,
P.O. Box 644, Bilbao, E-48080, Spain}
\address{$^4$EHU Quantum Center (EHUQC),
University of the Basque Country UPV/EHU,
P.O. Box 644, Bilbao, E-48080, Spain}
\address{$^5$Institute for Advanced Study,
Technical University of Munich,
Lichtenbergstra{\ss}e 2 a, Garching, G-85748, Germany}

\email{\href{mailto:mate.mihaly@wigner.hu}{mate.mihaly@wigner.hu}}
\email{\href{mailto:petrov.klara@wigner.hu}{petrov.klara@wigner.hu}}
\email{\href{mailto:szalay.szilard@wigner.hu}{szalay.szilard@wigner.hu}}
\email{\href{mailto:legeza.ors@wigner.hu}{legeza.ors@wigner.hu}}

\date{April 25, 2022}

\begin{abstract}
In this work, we present a brief overview of the fermionic mode optimization 
within the framework of tensor network state {methods} [PRL\textbf{117},210402(2016)],
and demonstrate that it has the potential 
to compress the multireference character of the wave functions
after finding optimal molecular orbitals (modes), based on entanglement minimization.
Numerical simulations have been performed
for the nitrogen dimer in the cc-pVDZ basis for the equilibrium and for stretched geometries.
\end{abstract}


\maketitle

\section{Introduction}
\label{sec:intro}

In the past two decades, singular value decomposition (SVD) based \emph{tensor network state} (TNS) methods have become vital alternative approaches to treat strongly correlated, i.e., multireference problems in quantum chemistry~\cite{White-1999,Legeza-2008,Chan-2008,Yanai-2009,Marti-2010c,Wouters-2014a,Legeza-2014,Szalay-2015a,Chan-2016,Baiardi-2020,Cheng-2022}.
These provide an approximation of an eigenstate of the ab inito Hamiltonian
as a product of low rank matrices or tensors,
thus the computational demands are governed by the ranks of the component tensors,
also known as \emph{bond dimension}.
In the course of the optimization procedure, the ranks of the tensors can be kept fixed, i.e., the optimization is carried out on a fixed submanifold, or, alternatively, they can be adapted dynamically to fulfill an a priory set error margin~\cite{Legeza-2003a,Holtz-2012a,Holtz-2012b}. In the latter case, the maximum rank depends strongly on the network topology~\cite{Legeza-2003b,Nakatani-2013,Murg-2015} and on the properties of the component tensors~\cite{Gunst-2018}. 
The optimal \emph{choice of the modes} (i.e., orbitals in quantum chemistry) is another key aspect that influences the efficiency of the TNS methods~\cite{Rissler-2006,Murg-2010a,Stein-2016},
because optimal modes can lead to localization of the correlation and entanglement in the system~\cite{Fertitta-2014}, and to a drastic reduction of the bond dimension~\cite{Krumnow-2016,Krumnow-2021}.
Therefore, a joint optimization strategy that optimizes both the tensors and the modes simultaneously
is expected to lead to a black box application of TNS methods for strongly correlated multireference problems.

The optimization of the orbitals is not a new concept,
\emph{localized molecular orbitals} (LMO) has a long history in quantum chemistry.
The aim of the localization of orbitals is twofold.
On the one hand, localization leads to chemically intuitive orbitals
for rationalizing electronic structure of molecular systems.
On the other hand, LMOs has proven to be useful
in making the high-level correlated quantum chemical methods more tractable computationally.
These methods are based on specially constructed unitary operators,
and usually involve the optimization of the expectation value of a specific operator.
Among many others, we can recall 
Foster--Boys localization~\cite{Foster-1960,Boys-1960},
which minimizes the radial extent of the localized orbitals,
or Pipek--Mezey localization~\cite{Pipek-1989},
which is based on maximizing the charge of each orbital.

In this work, we present a brief overview of  
the main aspects of the \emph{orbital optimization} protocol,
which is the quantum chemical application of the more general \emph{fermionic mode transformation}~\cite{Krumnow-2016,Krumnow-2021},
and 
demonstrate that it has the potential 
to compress the multireference character of the wave functions,
after finding optimal MOs, based on entanglement localization.
Numerical simulations are performed for the nitrogen dimer
for the equilibrium and for stretched geometries in the \mbox{cc-pVDZ} basis, 
which is a common basis for benchmark computations, developed by Dunning and co-workers~\cite{Dunning-1989}.
Note that we use the term \emph{``basis''} only for the atomic basis set cc-pVDZ.
From this, the \emph{``canonical MOs''} are obtained by the Hartree--Fock SCF optimization,
which form the \emph{``initial MOs''} of the DMRG calculation.
From this, the \emph{``optimized MOs''} are obtained by the orbital optimization,
which is our main concern.
The more general term \emph{``modes''} is used for orbitals only where the general aspects of the theory are emphasized.

The organization of this work is as follows: 
in Section~\ref{sec:theory}, we briefly recall the basics of matrix product states (MPS, a special case of TNS)
and orbital optimization;
in Section~\ref{sec:numerics}, we describe the numerical procedure applied,
and present our numerical results;
in Section~\ref{sec:conclusions}, we draw the conclusions.

\section{Theoretical background}
\label{sec:theory}

In this section, we present a brief overview of the joint optimization procedure
based on MPS and orbital optimizations,
while a more detailed description can be found in the original works~\cite{Murg-2010a,Krumnow-2016}.

In the context of nonrelativistic quantum chemistry,
the Hamiltonian takes the second quantized form
\begin{equation}
\label{eq:Hamiltonian}
H = \sum_{i,j=1}^d \sum_{\sigma\in\{\downarrow, \uparrow\}} 
t_{i,j} c_{i,\sigma}^\dag c_{j,\sigma}
+ \sum_{i,j,k,l=1}^d \sum_{\sigma,\sigma'\in\{\downarrow, \uparrow\}}
v_{i,j,k,l} c_{i,\sigma}^\dag c_{j,\sigma'}^\dag c_{l,\sigma'} c_{k,\sigma},
\end{equation} 
where $i,j,k,l$ are spatial indices and $\sigma,\sigma'$ are spin indices,
and $c_{i,\sigma}$ are the fermionic annihilation operators,
satisfying the canonical anti-commutation relations 
$\{c_{i,\sigma},c_{j,\sigma'}\}=0$ and 
$\{c_{i,\sigma}^\dag,c_{j,\sigma'}\}=\delta_{i,j}\delta_{\sigma,\sigma'}$.  
There are constraints among the elements of the spin-independent integrals $t$ and $v$ such that $H$ is Hermitian. 
The Hilbert space of $N$ interacting electrons in $d$ orbitals
is the $N$-electron subspace of the fermionic Fock space $\mathcal{F}_d \cong \bigotimes_{i=1}^d \mathcal{H}_i$,
which can be described by the basis constituted by all Slater determinants 
$\{ \ket{\alpha_1,\dots,\alpha_d}=\bigotimes_{i=1}^d \ket{\alpha_i} \}$,
where the occupation indices are $\alpha_i\in\{0,\downarrow,\uparrow,\downarrow\uparrow\}$, labeling
the basis in the occupation spaces $\mathcal{H}_i$ of orbitals $i\in\{1,2,\dots,d\}$.
The quantum many-body wave function can be written as a linear combination of all Slater determinants
\begin{equation}
\ket{\psi} = \sum_{\substack{\alpha_1,\ldots,\alpha_d\\\in\{0,\downarrow,\uparrow,\downarrow\uparrow\}}}
C_{\alpha_1,\ldots,\alpha_d}\ket{\alpha_1,\ldots,\alpha_d},
\label{eq:fulltensor}
\end{equation}
where the high-order coefficient tensor $C\in(\mathbb{C}^4)^{\otimes d}$ is determined
by the eigenvalue problem of the Hamiltonian given by Eq.~\eqref{eq:Hamiltonian}
in the $N$-electron subspace.
The full configuration interaction (full CI) wavefunction~\eqref{eq:fulltensor} can be
expressed as a linear combination of wave functions
corresponding to different excitation levels with respect to a reference determinant,
\begin{equation} 
\ket{\psi} = \sum_I C_I\ket{\psi_I},
\label{eq:ci}
\end{equation}
where the $I=0$ term $\ket{\psi_0}$ refers to the reference determinant, 
and the $I=1,2,3\dots$ terms $\ket{\psi_1},\ket{\psi_2},\ket{\psi_3}\dots$ refer
to the single, double, triple\dots excitations, respectively.
The coefficients $C_0,C_1,C_2,C_3\dots$ normalize the CI terms~\cite{Helgaker-2000}.

In the MPS representation, the wave function takes the form
\begin{equation}
\ket{\psi} =
\sum_{\substack{\alpha_1,\ldots,\alpha_d\\\in\{0,\downarrow,\uparrow,\downarrow\uparrow\}}} A^{\alpha_1}_{[1]}\cdots A^{\alpha_d}_{[d]}\ket{\alpha_1,\dots,\alpha_d},
\label{eq:DefinitionMPS}
\end{equation}
where the \emph{component tensors} are $A_{[i]}^{\alpha_i}\in \mathbb{C}^{D_{i-1}\times D_i}$,
with \emph{bond dimensions} $D_i$, and $D_0 = D_d = 1$.
Every state vector can be written in an MPS form
by applying consecutive SVDs~\cite{Vidal-2003b},
using sufficiently large bond dimensions,
however, this scales exponentially with $d$ in the generic case.
The restriction of the bond dimensions to a fixed value $D$
restricts the full state space to a sub-manifold.
We can then approximate an eigenstate of the Hamiltonian~\eqref{eq:Hamiltonian}
within this sub-manifold by the use of
the \emph{density matrix renormalization group} (DMRG) algorithm,
which, being an alternating least square method, optimises the entries
of the MPS tensors $A_{[i]}$ iteratively~\cite{Ostlund-1995,Verstraete-2004a,Verstraete-2004b, Legeza-2014,Szalay-2015a},
leading to a variational treatment of the eigenvalue problem of the Hamiltonian~\eqref{eq:Hamiltonian}.

Utilizing a unitary orbital-transformation $U \in \mathrm{U}(d)$,
a linear transformation of a set of fermionic annihilation operators $\{c_{i,\sigma}\}$ to a new set $\{d_{i,\sigma}\}$ satisfying the canonical anti-commu\-ta\-tion
relations can be obtained, i.e., $c_{i,\sigma} = \sum_{j=1}^d U_{i,j,\sigma} d_{j,\sigma}$. 
We note that in the presented system
it is not necessary to use different unitaries for spin up and down, $U_{i,j,\uparrow}=U_{i,j,\downarrow}$,
however, the implementation is applicable for the unrestricted case.
Under this transformation, 
the representation $G(U)$ can also be given on the Fock space~\cite{Krumnow-2016},
by which a fermionic wave function
$\ket{\psi(\id)}$ transforms to $\ket{\psi(U)} = G(U)^\dagger\ket{\psi(\id)}$
and the Hamiltonian written in terms of the transformed orbitals 
by $H(U) = G(U)^\dagger H G(U)$.
In the course of the DMRG algorithm,
the unitary $U$ is constructed iteratively from two-orbital unitary operators
by sweeping through the network.
At each micro-iteration step, the half-R\'enyi block entropy
$S_{1/2}(\rho_{\{1,2,\dots,k\}}) = 2\ln(\Tr\sqrt{\rho_{\{1,2,\dots,k\}}})$
is minimized by a two-orbital rotation.
(Here $\rho_{\{1,2,\dots,k\}}$ is the density operator of the first $k$ orbitals~\cite{Szalay-2021,Boguslawski-2013}.)
In practice, when turn to numerical simulation including orbital optimization,
it is favourable not to transform the operators themselves to keep robustness.
Rather it is practical to perform orbital optimization
in terms of the parameters $t$ and $v$ in the Hamiltonian.

At the end of the last DMRG sweep, 
the \emph{one-orbital entropies} $s_i$, 
the two-orbital \emph{mutual informations} $I_{i,j}:=s_i+s_j-s_{i,j}$,
the \emph{total correlation} $I_\text{tot} = \sum_i s_i$,
the \emph{correlation distance} $I_\text{dist} = \sum_{i,j} I_{i,j} \vert i - j\vert^2$,
the \emph{one-particle reduced density matrix} $\gamma_{i,j} = \langle c^\dagger_j c_i \rangle$,
and the \emph{occupation number distribution} $\langle n_i\rangle$ 
are calculated (where $i,j\in\{1,\ldots,d\}$).
Here, $s_i= -\Tr (\rho_i \ln \rho_i)$ and $s_{i,j}= -\Tr(\rho_{i,j}\ln\rho_{i,j})$
are the von-Neumann entropies of the \emph{one- and two-orbital reduced density operators} $\rho_i$ and $\rho_{i,j}$~\cite{Szalay-2021}.
The mutual information and more general correlation measures~\cite{Szalay-2015b,Szalay-2019} 
can be used not only for the optimization but also for the characterization of the chemical properties~\cite{Szalay-2017,Brandejs-2019}.
On the other hand, the eigenvalues and eigenvectors of the one-particle reduced density matrix
$\gamma$ define the \emph{natural occupation numbers} $\lambda_i$, and the \emph{natural orbitals} (NO).
An optimized ordering of orbitals along the
tensor network is calculated from the mutual informations $I_{i,j}$, using the Fiedler vector approach~\cite{Barcza-2011};
a new {complete active space vector} is calculated from the entropies $s_i$ for the {dynamically extended active space} (DEAS) procedure~\cite{Legeza-2003b};
and a new Hartree--Fock (HF) reference configuration is calculated from the occupations $\langle n_i\rangle$. 
These, together with the final rotated interaction matrices, are all used
as inputs for the subsequent orbital transformation macro-iteration.

\section{Numerical approach}
\label{sec:numerics}

In the numerical procedure,
the calculations are carried out 
for the nitrogen dimer in the cc-pVDZ basis~\cite{Dunning-1989}
for various bond lengths.
Systematically increasing the bond dimension $D_\text{opt}=16,64,256,512,1024,2048,4096$,
we have used nine DMRG sweeps, 
$20$ orbital optimization macro-iterations and
utilized $\mathrm{U}(1)$ symmetries only.
When $D_\text{opt}\leq512$ is used, after convergence is reached, 
large scale DMRG calculations are performed with bond dimension $D=4096$.
We note that, for large $D_\text{opt}$ bond dimension, orbital optimization converges already after 4-5 macro-iterations.

\begin{figure}
\centerline{
  \includegraphics[width=0.333\textwidth]{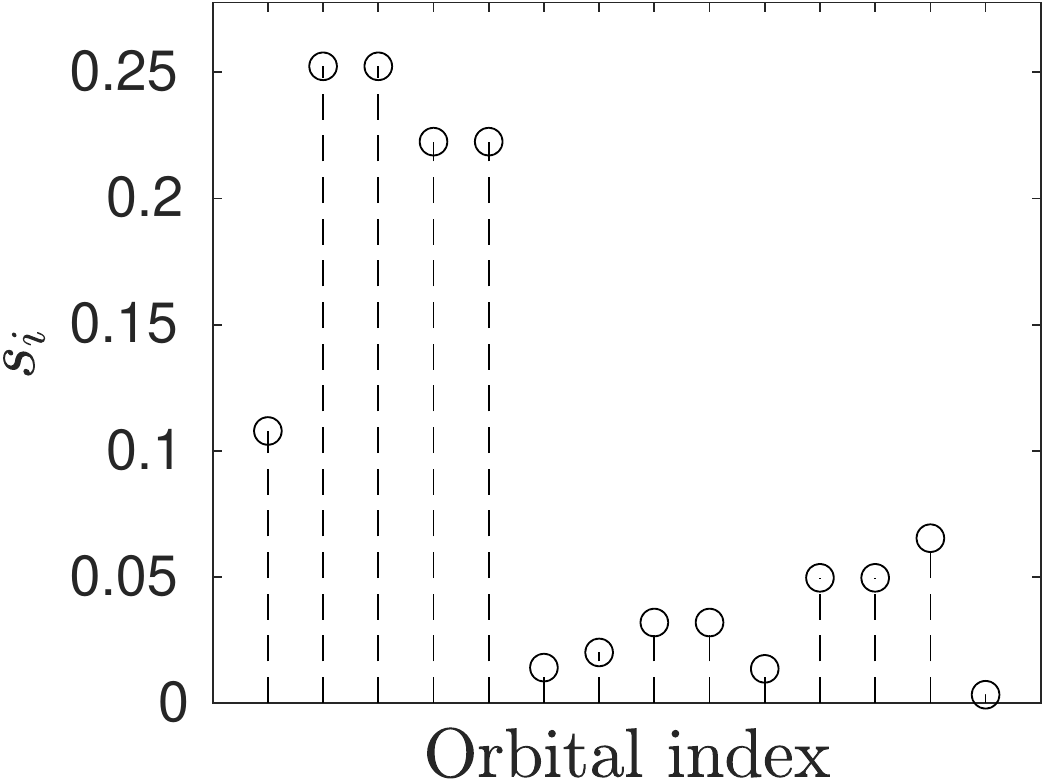}
  \includegraphics[width=0.333\textwidth]{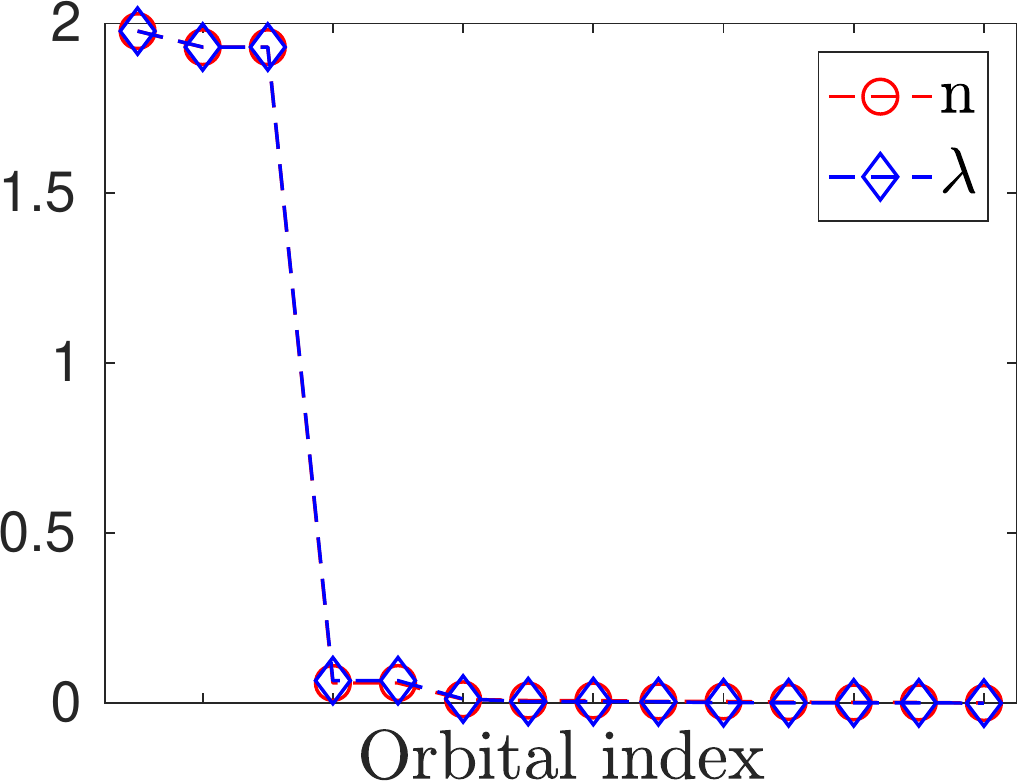}
  \includegraphics[width=0.36\textwidth]{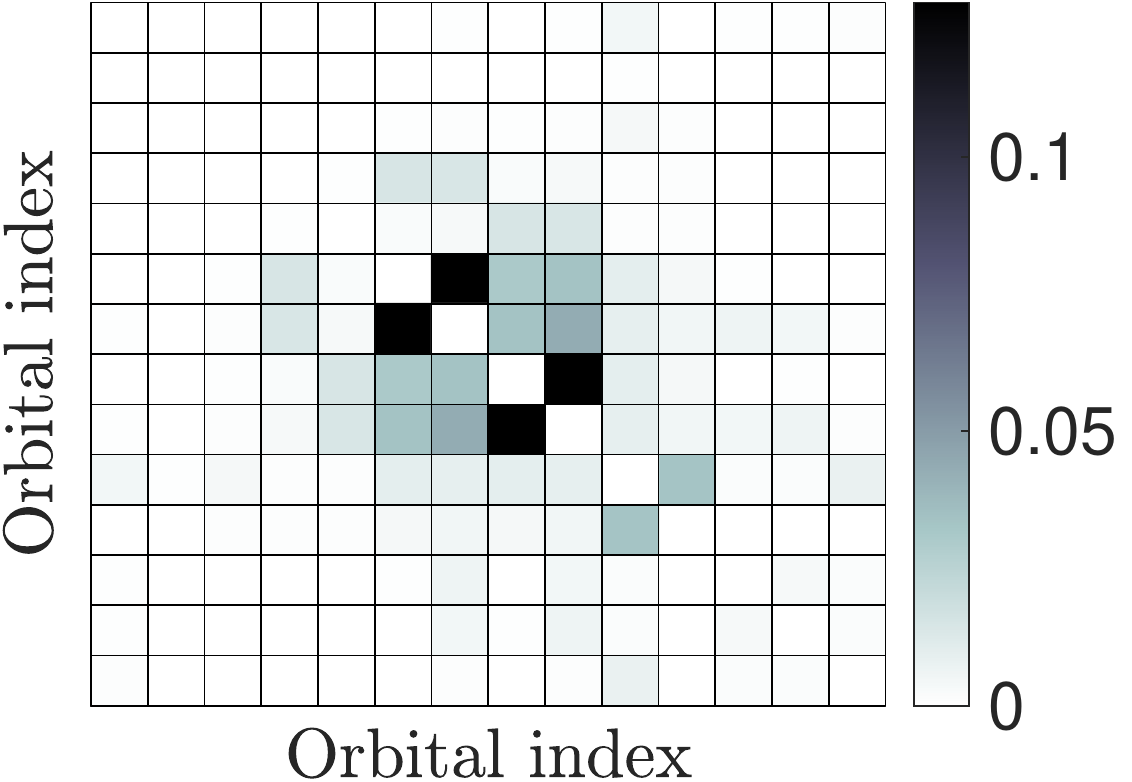}
}
\centerline{
\includegraphics[width=0.333\textwidth]{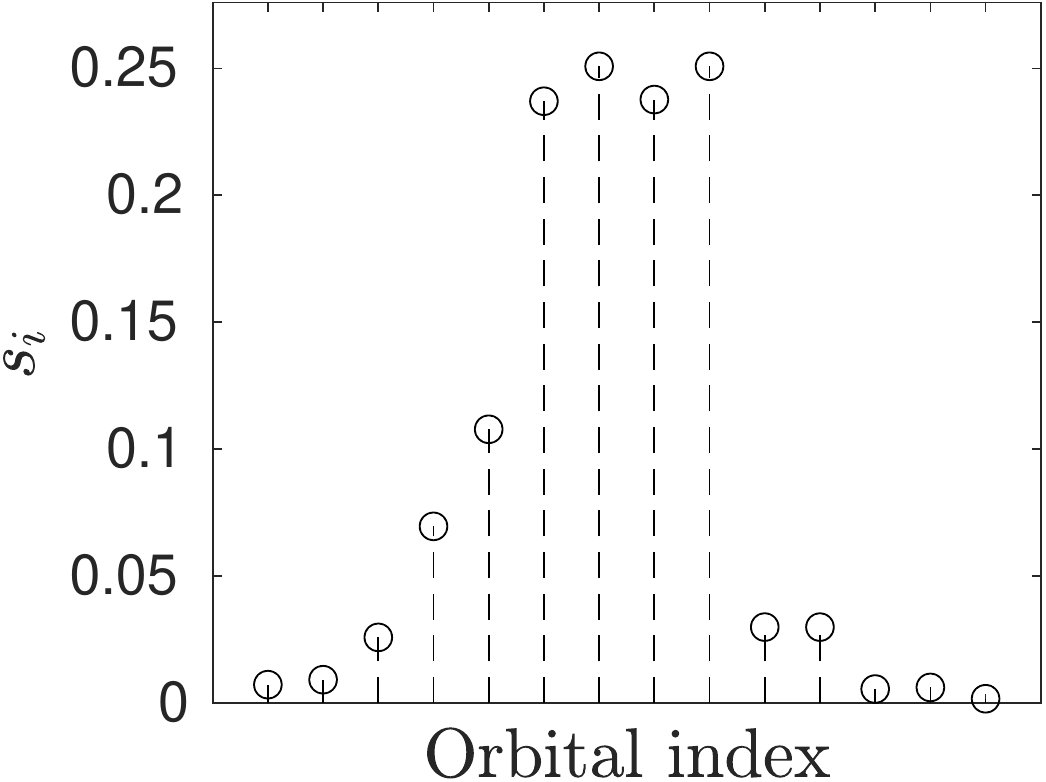}
  \includegraphics[width=0.333\textwidth]{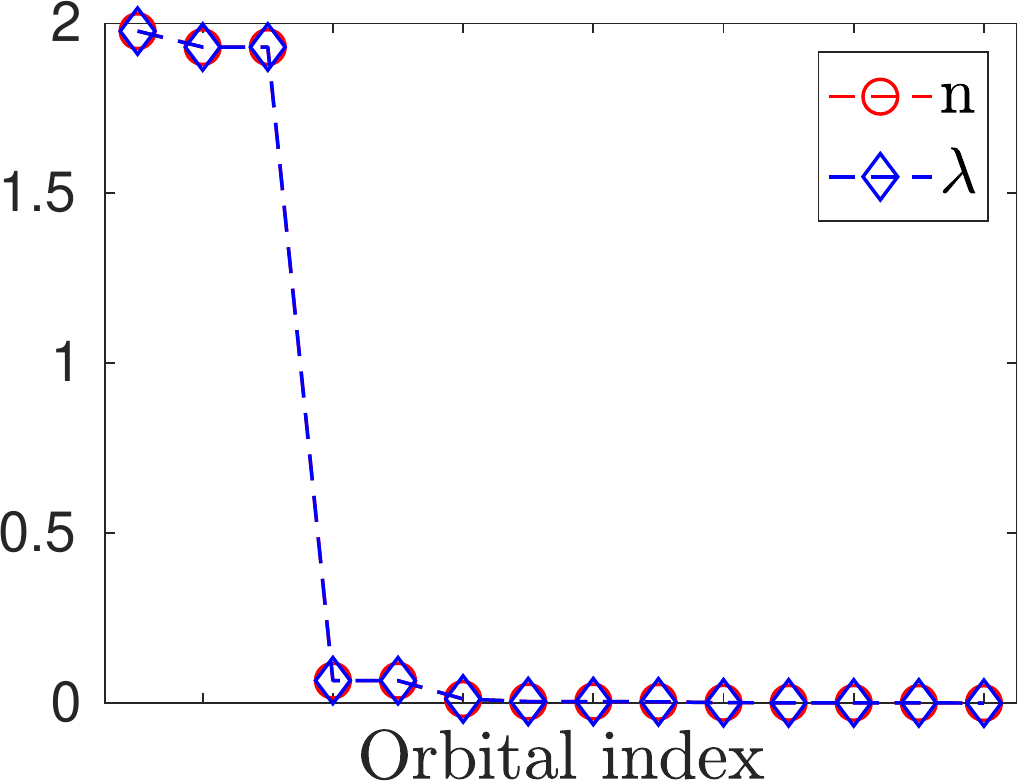}
  \includegraphics[width=0.36\textwidth]{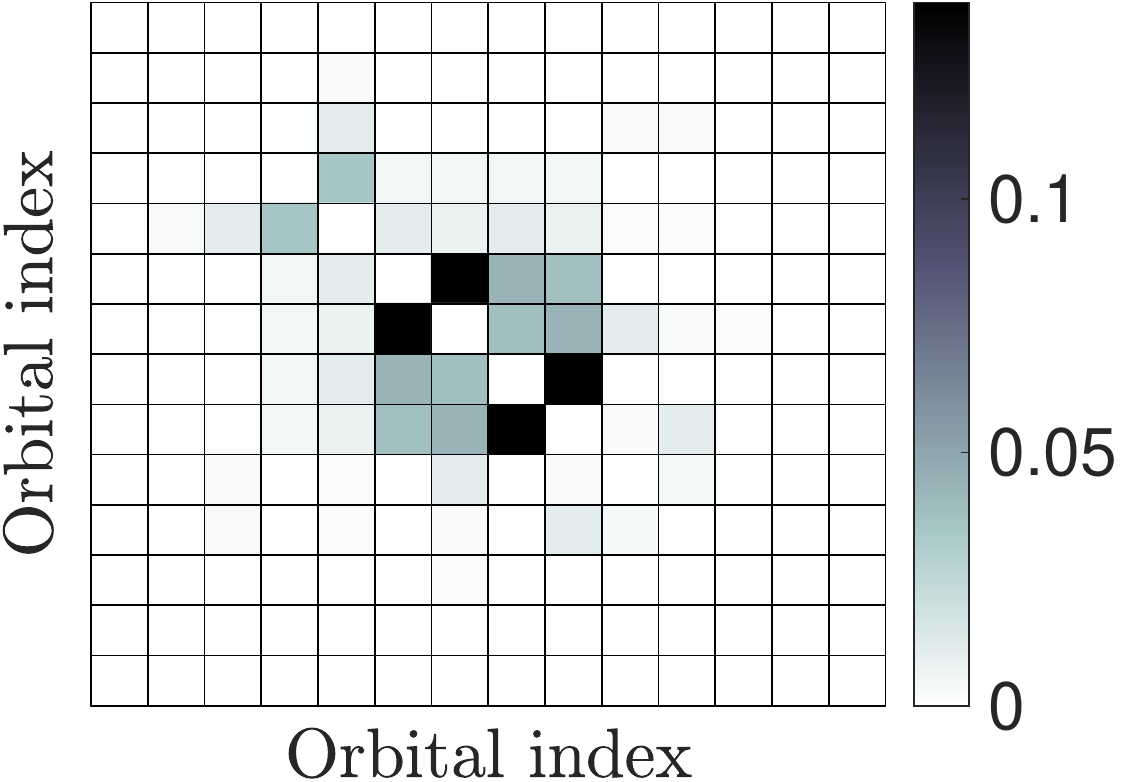}
}
\caption{Orbital entropy profiles $\{s_i\}$, sorted values of the natural orbital occupation numbers $\{\lambda_i\}$, 
and occupation numbers $\{\langle n_i\rangle\}$,
and two-orbital mutual informations $\{I_{i,j}\}$ for the 
initial MOs in CAS(6,14) (first row), and after the $20^\text{th}$ orbital optimization macro-iterations (second row) for the nitrogen dimer for bond length $r=2.118a_0$ using bond dimension $D=4096$.}
\label{fig:entropies-r2118}
\end{figure}

In earlier works, it has been demonstrated that DMRG
calculations after orbital optimization can lead to
significantly more accurate results for the same
computational complexity (e.g., same truncated bond dimension); or
to much lower truncated bond dimensions, needed to reach the same accuracy,
due to the tremendous reduction of the entanglement in the system~\cite{Krumnow-2016,Krumnow-2021}.
Here we focus on the emerging MOs and on the structure of the wave function.
Therefore, first we choose a small active space, namely, $6$ electrons on $14$ orbitals, CAS(6,14),
for which calculations can be performed in the full-CI limit.
Here, a very large value of $D_\text{opt}=4096$ is enforced.
A selected set of quantities,
based on concepts of quantum information theory,
to monitor numerically the performance of the fermionic orbital optimization
are shown in Fig.~\ref{fig:entropies-r2118},
for the equilibrium geometry with bond length $r=2.118a_0$,
obtained in the initial MOs (first row),
and for the optimized MOs (second row).
Here it can clearly be seen that the orbital optimization has no effect,
except that the MOs have been reordered along the DMRG chain
in order to reduce the correlation distance in the system
(calculated from the two-orbital mutual informations plotted also in Fig.~\ref{fig:entropies-r2118}),
$I_\text{dist} = 18.8575$ changes to $12.6095$.
The ground state energy $E = -109.0931\mathrm{Ha}$
remains invariant under the action of the unitary group,
the single-orbital entropy profiles changed only marginally
(plotted also in Fig.~\ref{fig:entropies-r2118}),
so $I_\text{tot}$, (calculated from the single-orbital entropies)
changes slightly from $1.3373$ to $1.2696$.
Three MOs are almost doubly occupied, and,
although canonical MOs have been used,
$\langle n_i\rangle$ and $\lambda_i$ fall on the top of each other for the initial and optimized MOs, resembling the characteristics of NO-like orbitals. 
The sharp Fermi edge indicates that the system is
weakly correlated, i.e., a single-reference problem.

\begin{figure}
\centerline{
  \includegraphics[width=0.333\textwidth]{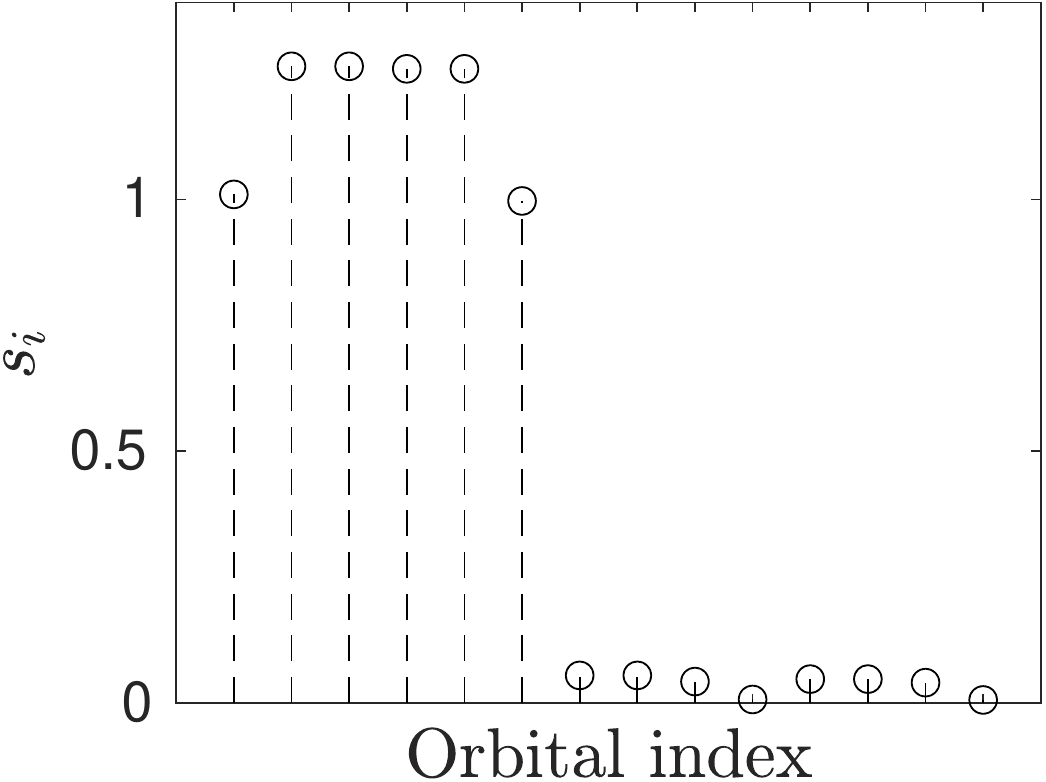}
  \includegraphics[width=0.333\textwidth]{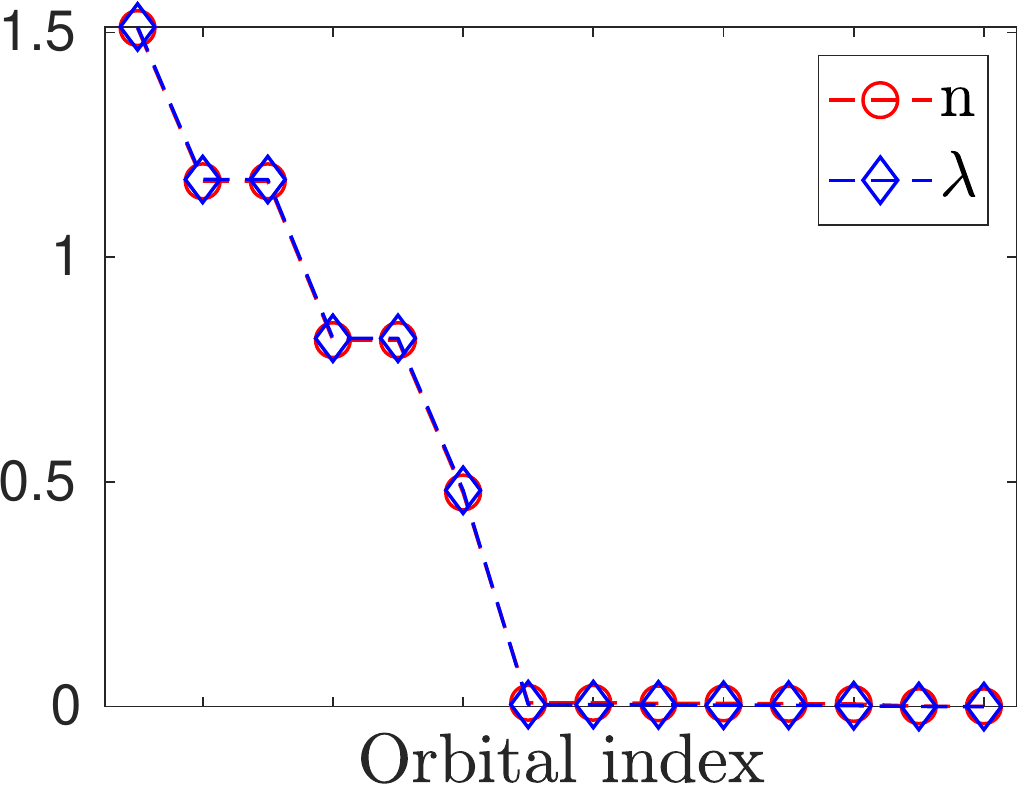}
  \includegraphics[width=0.35\textwidth]{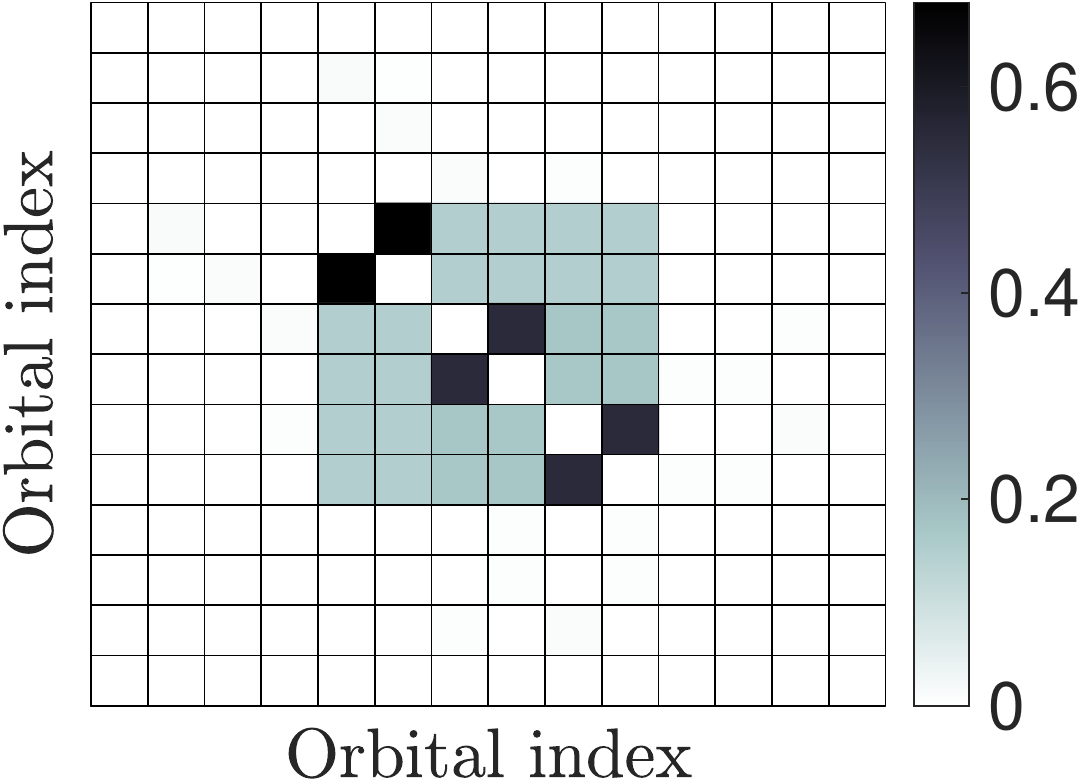}
}
\centerline{
\includegraphics[width=0.333\textwidth]{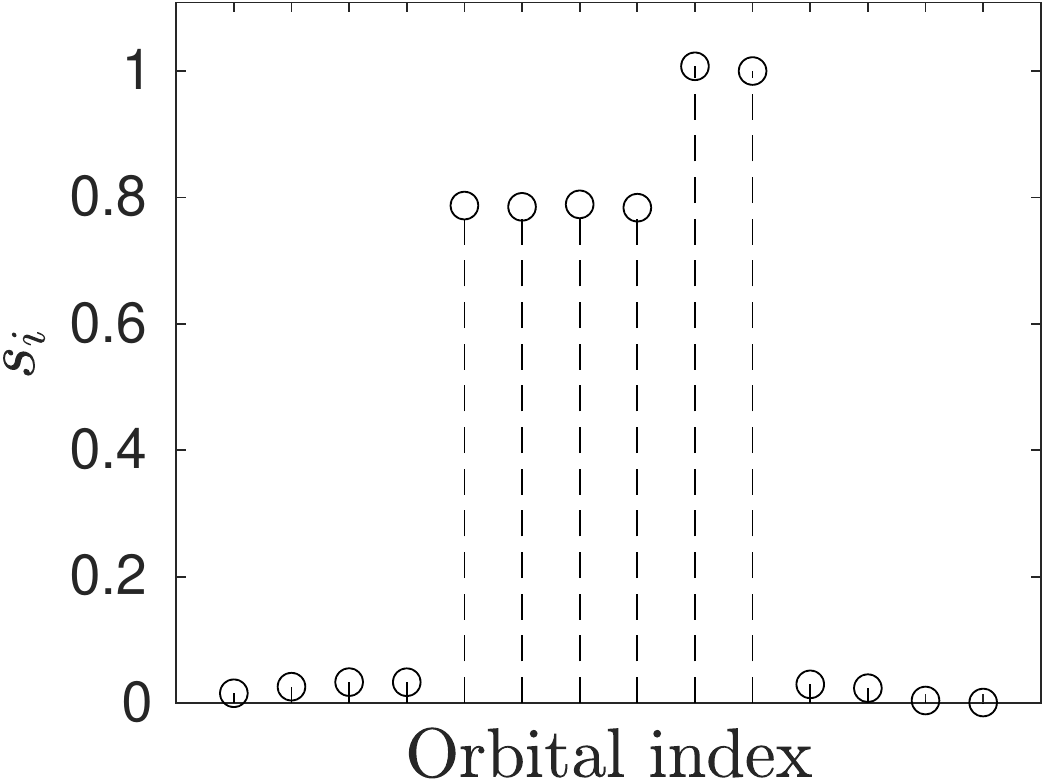}
  \includegraphics[width=0.333\textwidth]{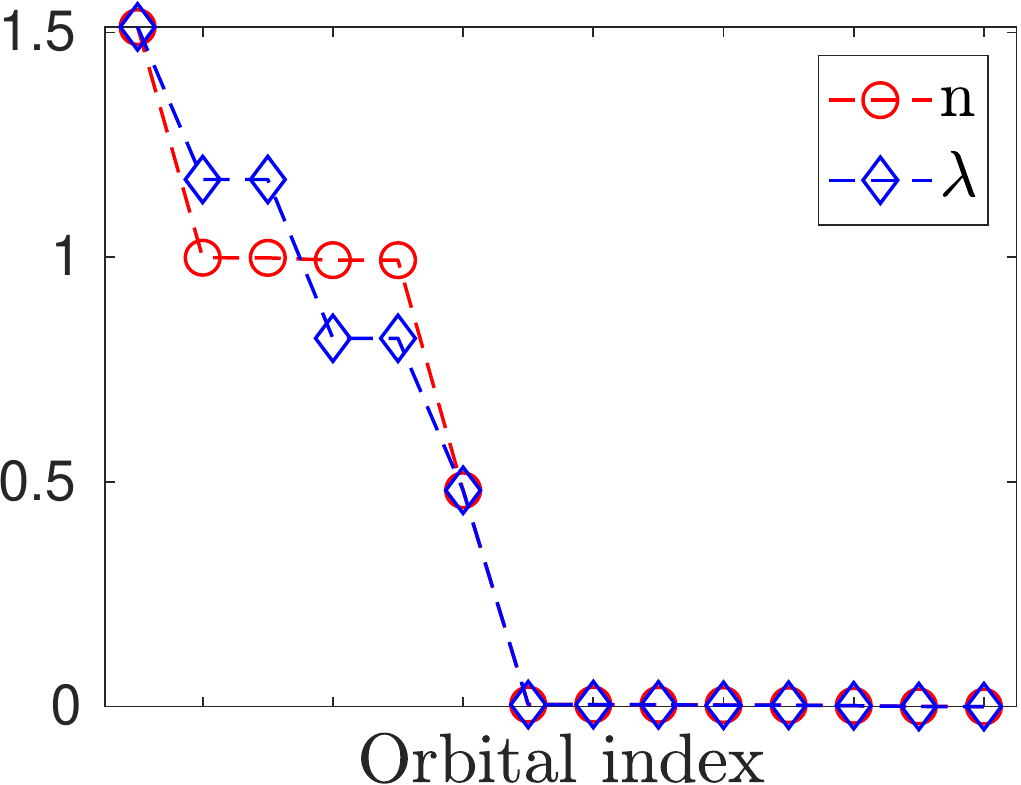}
  \includegraphics[width=0.35\textwidth]{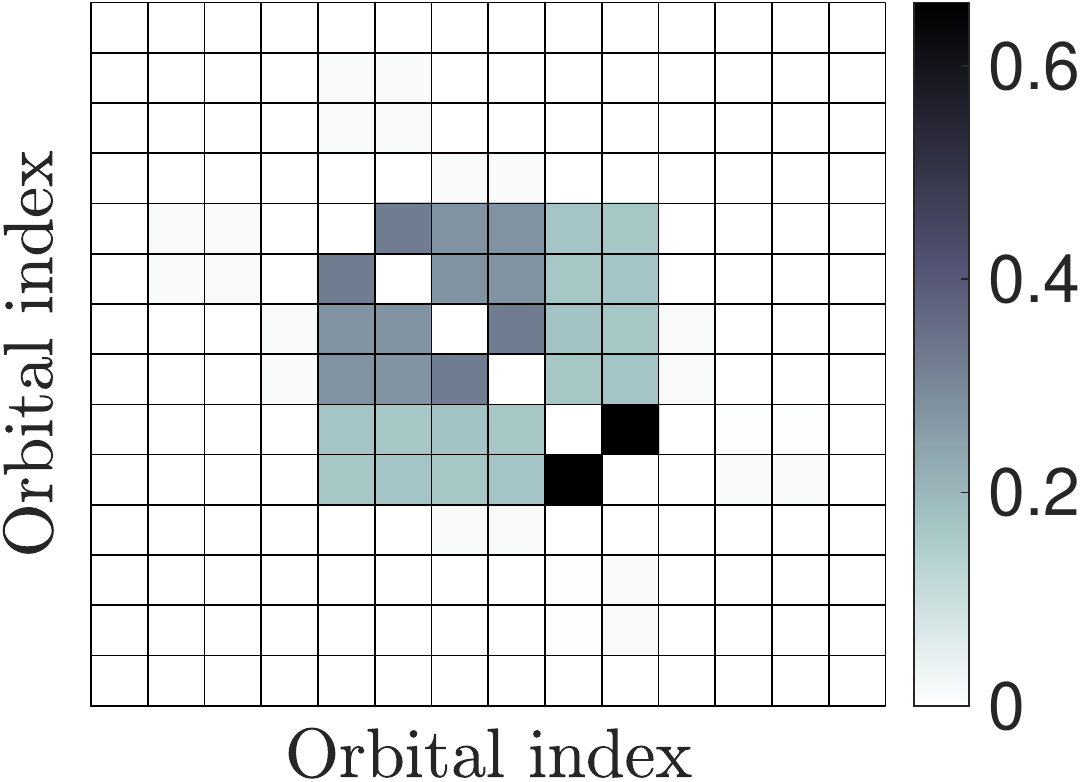}
}
\caption{(a) Similar to Fig.\ref{fig:entropies-r2118}, but for a stretched geometry at $r=4.200a_0$.}
\label{fig:entropies-r4200}
\end{figure}

In contrast to this, for a stretched geometry $r=4.200a_0$, the sharp drop off in $\langle n_i\rangle$ and $\lambda_i$ at the Fermi edge disappears, see in Fig.~\ref{fig:entropies-r4200}.
The corresponding six partially occupied orbitals possess very large orbital entropies,
indicating that these orbitals are in mixed states,
and are highly entangled with the rest of the system. 
The two orbitals with occupation number close to $1.5$ and $0.5$ are the $\sigma$ bonding and anti-bonding orbitals, while the four orbitals with $0.5 \leq \langle n_i\rangle \leq 1.5$ are orbitals with $\pi$ symmetry.
The underlying bond breaking effect
has already been analyzed in terms of entropies in Ref.~\cite{Boguslawski-2013}, however, such analysis depends on the choice of basis and on the modes optimized, as will be addressed below.
Carrying out MO optimization, new MOs are found, which are no longer NO-like
(see the difference between the profiles of $\langle n_i\rangle$ and $\lambda_i$).
Here, $\langle n_i\rangle=1$
and $s_i\simeq0.8$ for four orbitals, i.e.,
the electrons are uniformly distributed on the corresponding $\pi$ orbitals. The orbital entropy
for the two $\sigma$ orbitals remains close to unity, which also signals that the $\pi$ bonds break first.
Note that the results of this quantitative analysis in terms of orbital entropies
are the opposite as those in Ref.~\cite{Boguslawski-2013},
which demonstrates again that the entropic analysis is basis and mode transformation dependent.
Although the ground state energy does not change during the macro-iterations, $E=-108.7935\mathrm{Ha}$, 
the orbital entropies are reduced.
Therefore, the overall quantum correlation encoded in the wave function, $I_\text{tot}$, 
reduces from $7.3585$ to $5.3188$, and
$I_\text{dist}$ from $36.6702$ to $28.3941$.
Further stretching the nitrogen dimer,
the orbital entropies of the partially occupied orbitals scale towards $\ln(4)\simeq 1.38$ in the initial MOs, 
while they are reduced to $\ln(2)\simeq 0.69$ in the optimized MOs.
For the optimized MOs, $\langle n_i\rangle$ takes values very close to one or zero,
i.e., the six electrons are distributed uniformly on the the six partially occupied orbitals.
Note that these six orbitals are almost uncorrelated with the rest of the orbitals, i.e., 
the problem reduces to a CAS(6,6), as expected.
In this almost half-filled configuration, 
the empty and doubly occupied configurations provides no contribution for the orbitals of CAS(6,6),
giving $s_i\simeq\{\ln(2),0\}$.
For $r=20.000a_0$, even the initial MOs lead to the latter configuration.

\begin{figure}
\centerline{
  \includegraphics[width=1.0\textwidth]{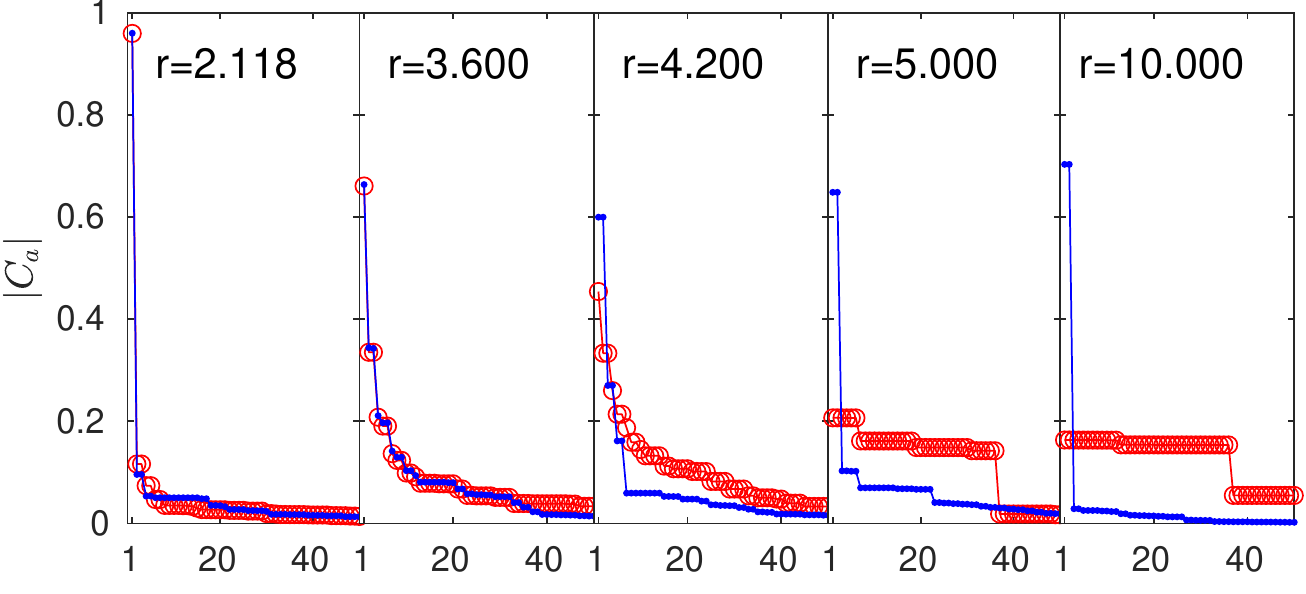}
}
\caption{Absolute value of the $50$ largest $C_a$ elements of the coefficient tensor in decreasing order
for various bond lengths for the initial MOs (red) and for the optimized MOs (blue),
obtained in the full-CI limit in CAS(6,14).
Here the relabelling $C_a$ for $a=1,2,\dots,4^d$ is used
for the $C_\alpha$ elements of the coefficient tensor,
such that $\vert C_a\vert \geq \vert C_b\vert$ if $a<b$.}
\label{fig:CI_6_14}
\end{figure}

Besides the entropic quantities, it is interesting to study 
the $C_\alpha$ entries of the coefficient tensor~\eqref{eq:fulltensor},
extracted from the MPS wave function obtained by the DMRG algorithm. 
Figure~\ref{fig:CI_6_14} shows the absolute value of the
$50$ largest $C_\alpha$ elements of the coefficient tensor
in decreasing order, for various bond lengths.
It is clearly visible that at the equilibrium geometry ($r=2.118a_0$) for the initial MOs (red),
there is one determinant of weight almost one,
and the remaining coefficients are smaller by at least an order of magnitude.
This single-reference property, however, changes,
as the nitrogen dimer is stretched,
and the leading coefficient gets smaller and smaller,
until degenerate plateaus appear.
This multireference behaviour is in accordance with the entropic analysis discussed above. 

When orbital optimization is also utilized,
the resulting profile of the $C_\alpha$ entries of the coefficient tensor changes significantly
with increasing bond length,
compared to the initial MOs,
as is shown in Fig.~\ref{fig:CI_6_14} by blue color.
For the equilibrium geometry at $r=2.118a_0$,
the difference between the initial and optimized MOs is negligible,
as the initial MOs already provides a single-reference approach for the problem.
In contrast to this, for $r\ge3.600a_0$,
the effect of orbital optimization becomes more drastic.
For $r\ge 4.200a_0$, the leading coefficients become two-fold degenerate,
corresponding to a
determinant and its spin flipped component,
and their weight increase rapidly to the saturation value of $1/\sqrt{2}$ 
with increasing bond length.
The plateau observed in the initial MOs for 
$r\ge10.000a_0$ completely disappears.
Along these lines,
the sum of the square of the absolute values of the largest CI coefficients
also shows a more rapid convergence to unity in the optimized modes for the stretched geometries,
as shown in Fig.~\ref{fig:sumCIsq}. 
Additionally, a more CI-based analysis is given in Appendix~\ref{sec:appCI}.
The fast decay of the $C_\alpha$ tensor coefficients in the optimized MOs leads to a more suitable basis for DMRG, thus lower computational demands are needed to reach the same level of accuracy,
as mentioned before.

\begin{figure}
\centerline{
    \includegraphics[width = 0.8 \textwidth]{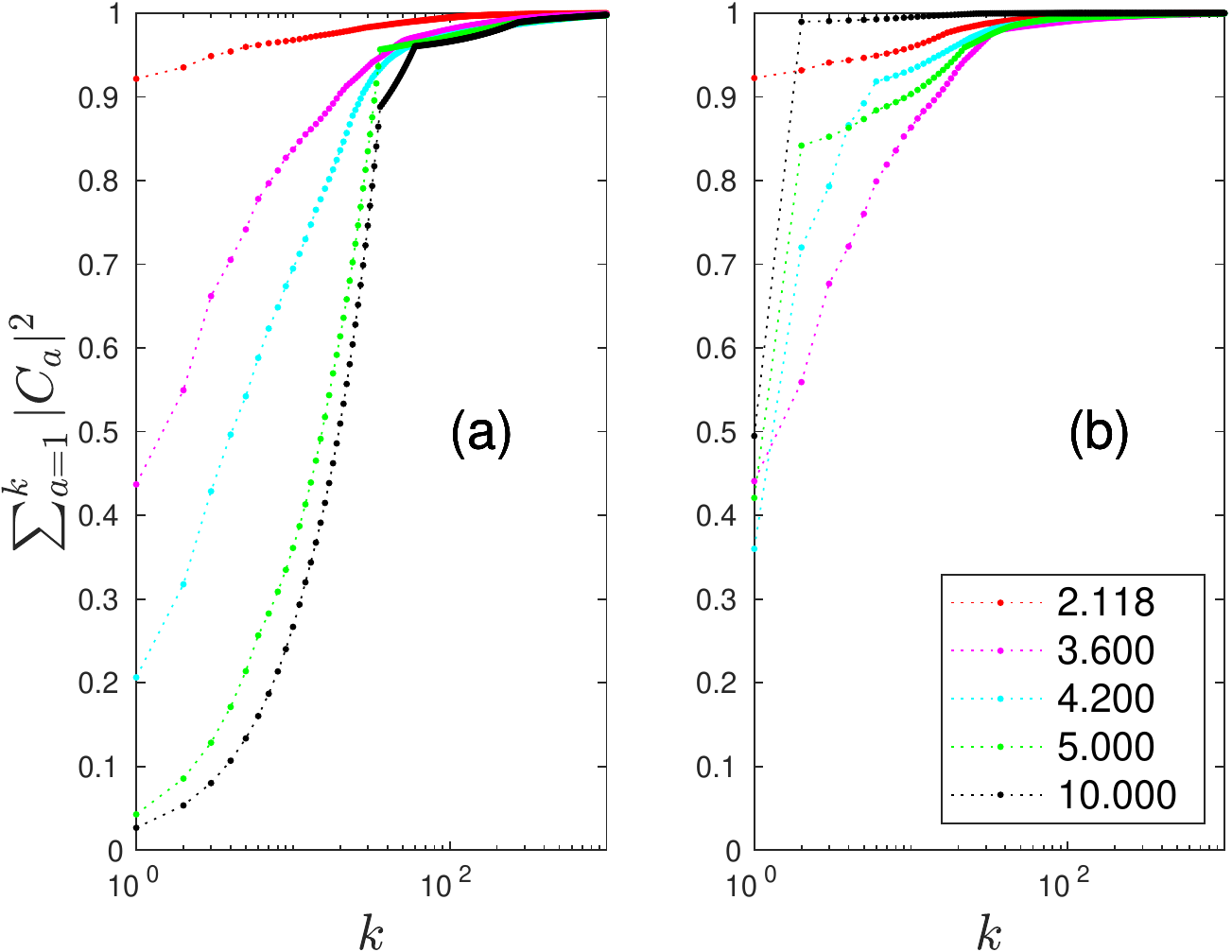}
}
\caption{(a) Sum of the square of the absolute values of the $1000$ largest CI coefficients
for the nitrogen dimer in the CAS(6,14) for various bond lengths, extracted from the MPS wave function, obtained by the DMRG algorithm, with a bond dimension $D=4096$. 
(b) Similar to (a), but for the optimized MOs.}
\label{fig:sumCIsq}
\end{figure}

Repeating the same analysis but for the full space, i.e., for CAS(14,28),
similar conclusions have been reached.
Here, however, the full-CI limit could not be reached,
thus the effects of bond dimension truncation also influence the results.
The corresponding entropy plots are summarized in Appendix~\ref{sec:appQIT},
while Fig.~\ref{fig:ci-orig}(a) shows the absolute value of the first $25$ largest $C_\alpha$ tensor coefficients
up to double excitation levels
in decreasing order for various bond lengths.
The reference determinant was obtained by the occupation number profile $\langle n_i \rangle$.
\begin{figure}
  \centerline{
  \includegraphics[width = 0.35 \textwidth]{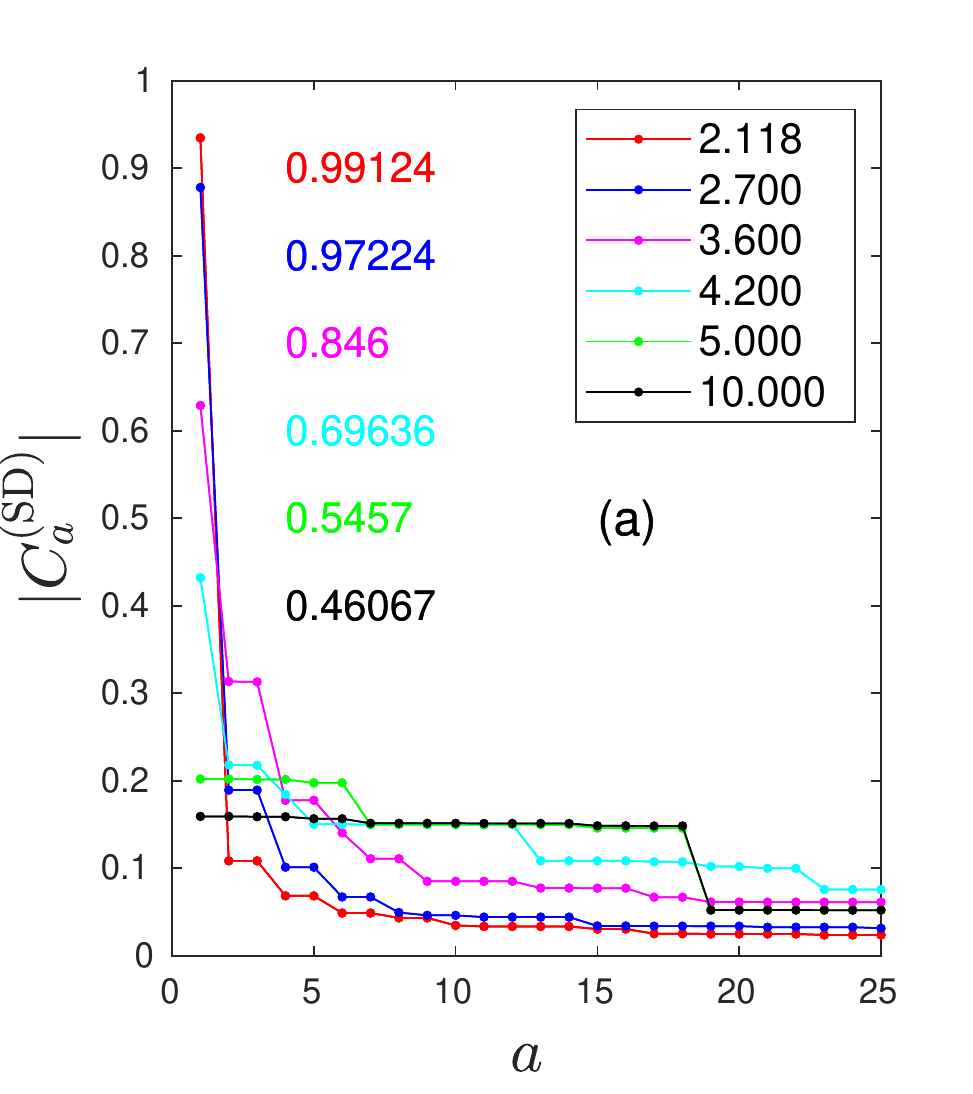}
  \hskip -0.45cm
  \includegraphics[width = 0.35 \textwidth]{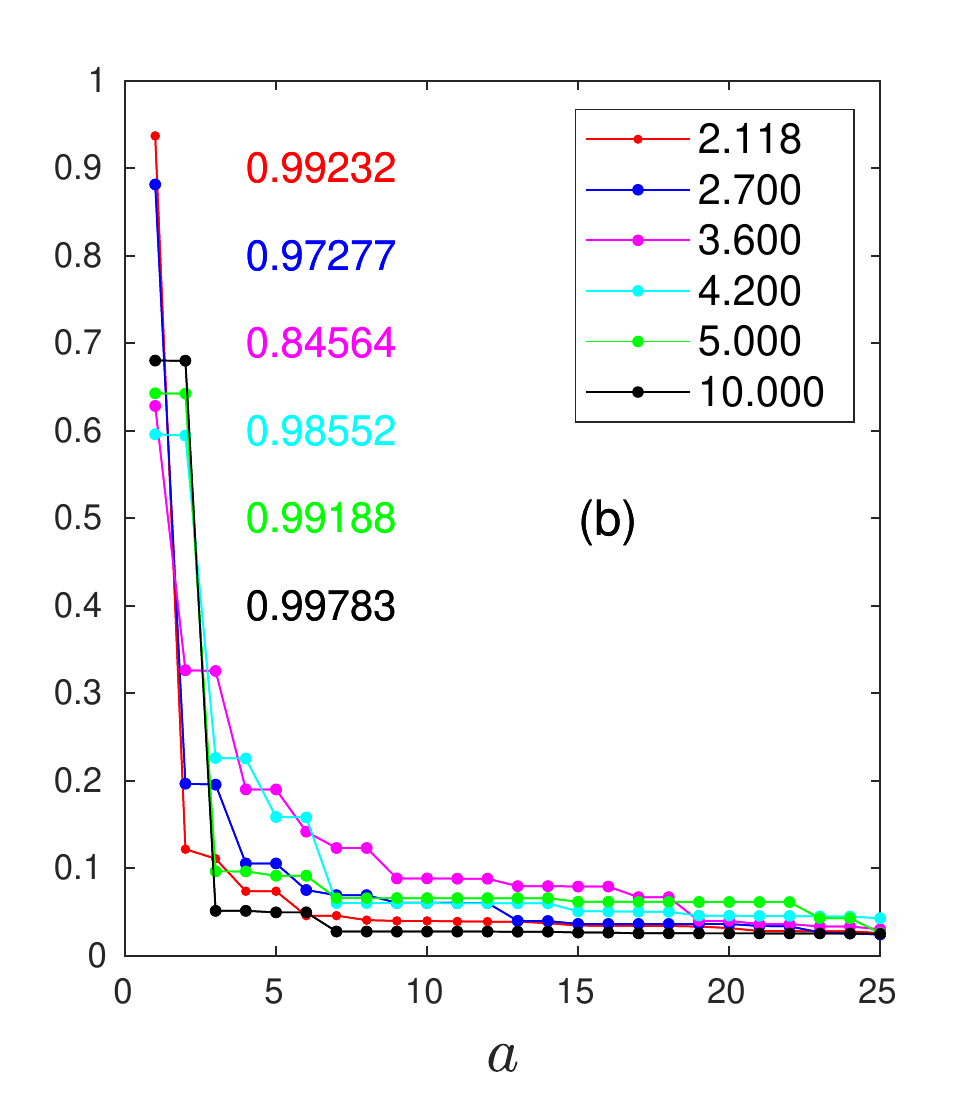}
  \hskip -0.45cm
  \includegraphics[width = 0.35 \textwidth]{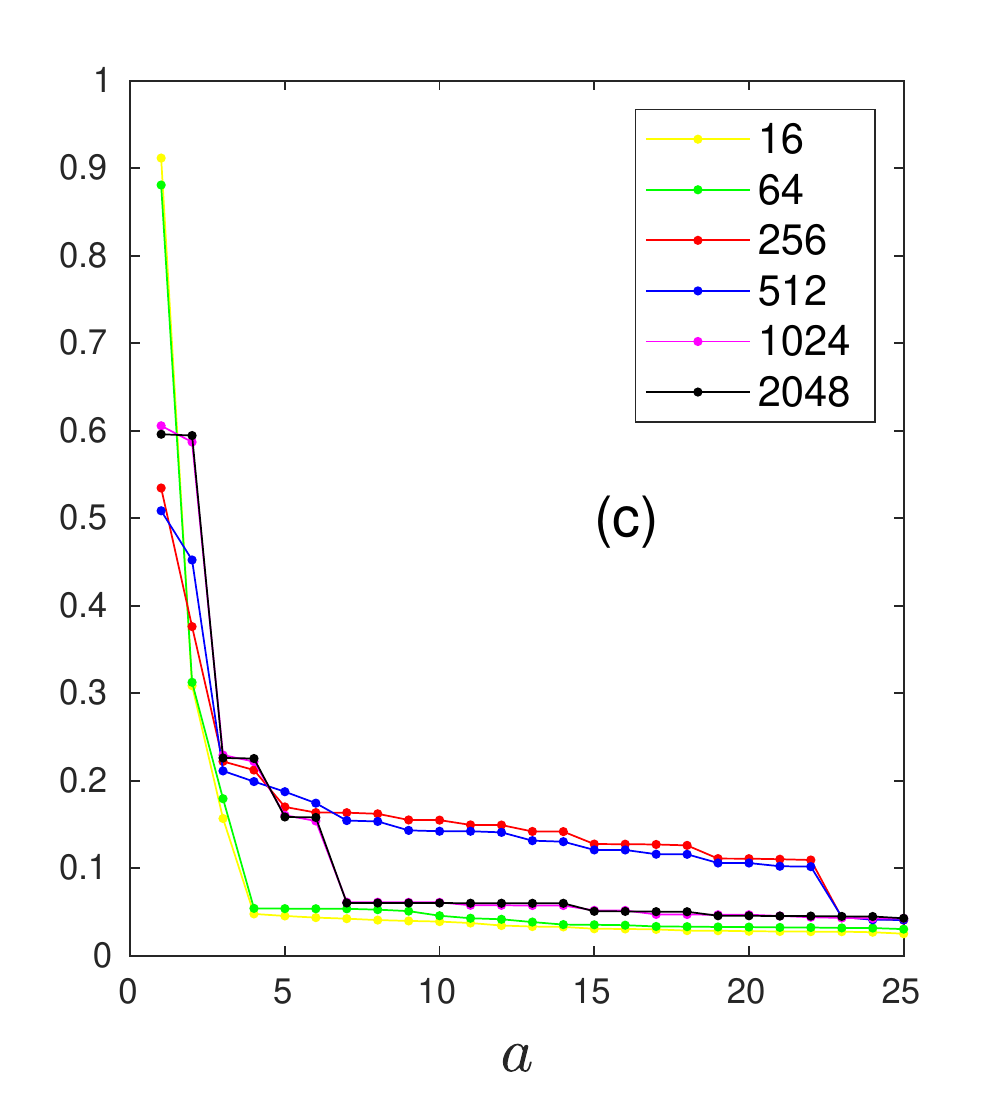}
}
\caption{(a) Absolute values of the first 25 largest CI coefficients including single and double excitation levels for the nitrogen dimer in the full space CAS(14,28) for various bond lengths, extracted from the MPS wave function, obtained by the DMRG algorithm, with a bond dimension $D=4096$. 
The inscribed numbers are the norm squares of the wave function component corresponding to single and double excitations  for the various bond lengths.
(b) Similar to (a), but for the optimized MOs with $D_\text{opt}=512$.
(c) Convergence of the absolute values of the first $25$ largest CI coefficients including single and double excitation levels for $r=4.200a_0$ for the optimized MOs, as a function of the bond dimension $D$.
}
\label{fig:ci-orig}
\end{figure}
Here, for the equilibrium bond length $r=2.118a_0$,
a sharp Fermi edge separates again the almost doubly occupied seven orbitals from the remaining virtual orbitals
for both the initial and the optimized MOs.
For the stretched geometries, six electrons get again shared among six orbitals,
which get superposed under the action of orbital optimization.
Since these orbitals are only marginally correlated with the rest of the orbitals,
the previous analysis holds for the corresponding CAS(6,6) space.

According to Fig.~\ref{fig:ci-orig}(a),
the absolute values of the leading $C_\alpha$ coefficients of the tensor,
together with the norm squares of the wave function components corresponding to single and double excitation levels
systematically decrease
with $r$ increasing from $2.118a_0$ to $10.000a_0$, as expected. 
Thus, in the large $r$ limit, higher level excitations
besides singles and doubles get more and more weight.
Here, we have used the original HF determinant as reference determinant,
in which configuration the first seven orbitals are doubly occupied.
For the optimized MOs, however, the leading coefficient increases drastically, 
and, again, the fast decay of the values is observed, see in Fig.~\ref{fig:ci-orig}(b).
The two-fold degeneracy of the leading coefficient, on the other hand, is sensitive to the bond dimension.
This is illustrated in Fig.~\ref{fig:ci-orig}(c) for $r=4.200$ for the optimized MOs obtained with $D_\text{opt}=512$,
as a function of $D$.
Here, two reference determinants, 
given by the occupation number profile,
connected by spin flip transformation,
have been used.
It is clear that for small bond dimensions, $16\leq D \leq 256$,
the problem looks like a single-reference one in the optimized MOs,
while for larger $D$ values, the correct degeneracy of the leading coefficient 
is recovered.
In addition, for $D=4096$, the norm square of the wave function component corresponding to single and double excitation levels
gets close to unity, see in Fig.~\ref{fig:ci-orig}(b), indicating that
orbital optimization has the potential to convert higher level excitations to lower ones, i.e., 
compressing multireference character of wave functions.
This provides a significantly more optimal MOs for DMRG computation,
which is also validated by the resulting lower energies for truncated bond dimensions.
Similar conclusion has also been drawn for the two-dimensional spinless fermionic lattice models 
commonly studied in solid-state physics,
where a single determinant is suitable for the description of the quantum many-body wave function
for the non-interacting case, and for infinitely large interaction~\cite{Krumnow-2021}.

\section{Conclusions}
\label{sec:conclusions}
In this work, we have presented a brief overview of the main aspects
of a joint optimization procedure for tensor network state methods,
when optimization is carried out on a fixed rank MPS manifold, and on the manifold of MOs.
Numerical illustrations were given
for the nitrogen dimer in the cc-pVDZ basis for the equilibrium and for stretched geometries.
We have analyzed the properties of the wave function,
based on various entropic quantities, and on the profile of the coefficient tensor, highlighting 
the basis and orbital transformation dependent nature of such quantities.
The corresponding method, dubbed orbital optimization, has the potential to reduce significantly
the correlation and entanglement encoded in the quantum many-body wave function, and 
to convert coefficients of higher level excitations to those of lower level ones,
resulting in a rapidly decaying entries of the wave function coefficient tensor.
These all together provide compression of the multirefernce character of wave functions, and
significantly more optimal MOs for TNS and conventional multireference methods.

\bigskip

\begin{center}
\textit{This work is dedicated to the memory of J\'anos Pipek.\\
His friendly personality and excellent lectures are remembered with great fondness.}
\end{center}

\section*{Acknowledgments}
This research has been supported by the Hungarian National Research,
Development and Innovation Office (NKFIH) through Grants Nos.~K120569 and K134983, 
and the ``Frontline'' Research Excellence Programme (Grant No.~KKP133827);
and by the Quantum Information National Laboratory of Hungary. 
\"O.L. acknowledges the Hans Fischer Senior Fellowship programme funded by the Technical University of Munich -- Institute for Advanced Study.
The development of DMRG libraries has been supported by the Center for Scalable and Predictive methods for Excitation and Correlated phenomena (SPEC), funded as part of the Computational Chemical Sciences Program by the U.S.~Department of Energy (DOE), Office of Science, Office of Basic Energy Sciences, Division of Chemical Sciences, Geosciences, and Biosciences at Pacific Northwest National Laboratory.

\vspace{60pt}

\appendix

\section{Quantities to monitor orbital basis and correlations}
\label{sec:appQIT}

\begin{figure}
\centerline{
  \includegraphics[width=0.333\textwidth]{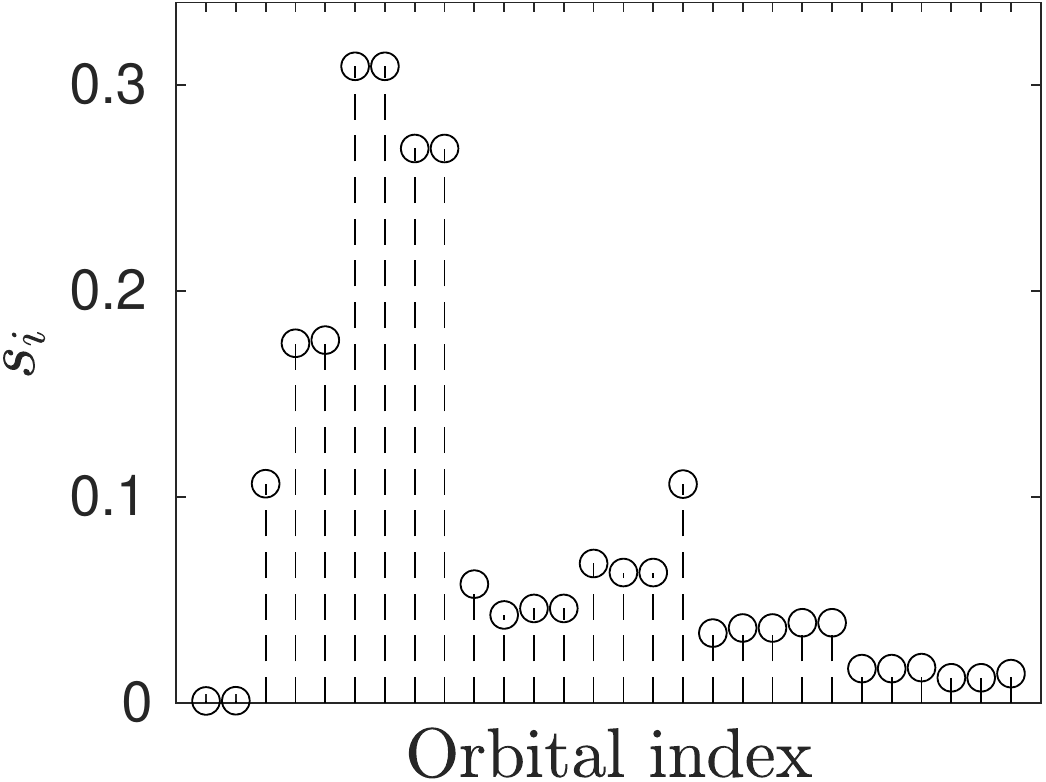}
  \includegraphics[width=0.333\textwidth]{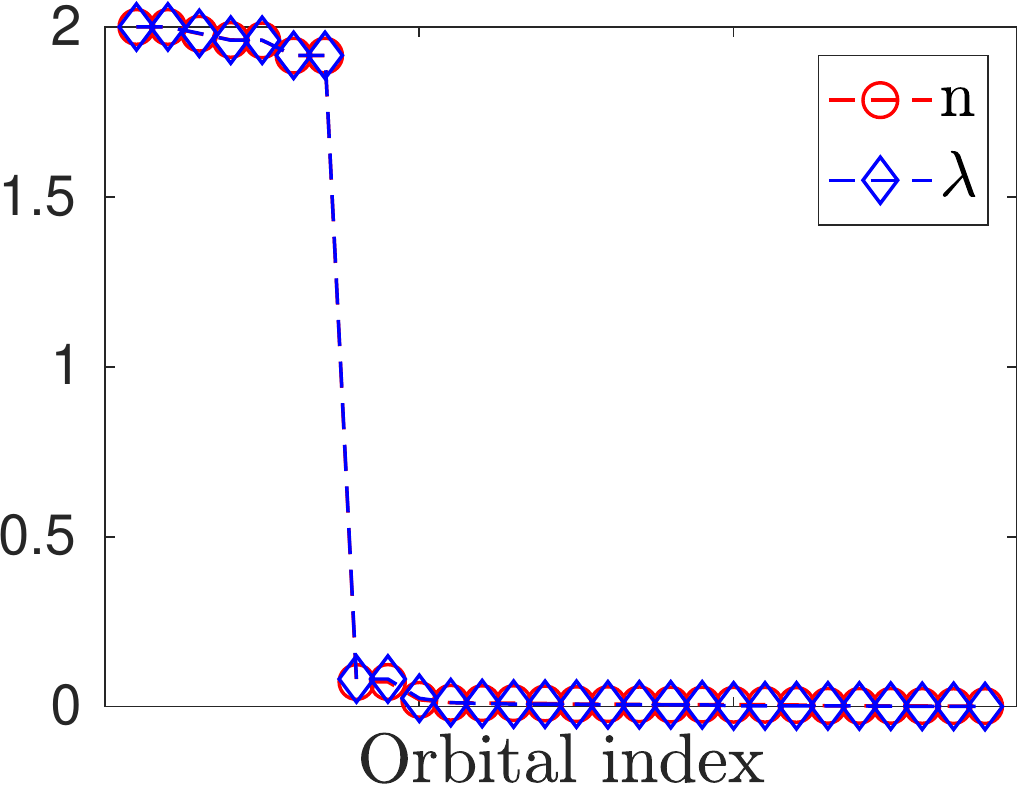}
  \includegraphics[width=0.35\textwidth]{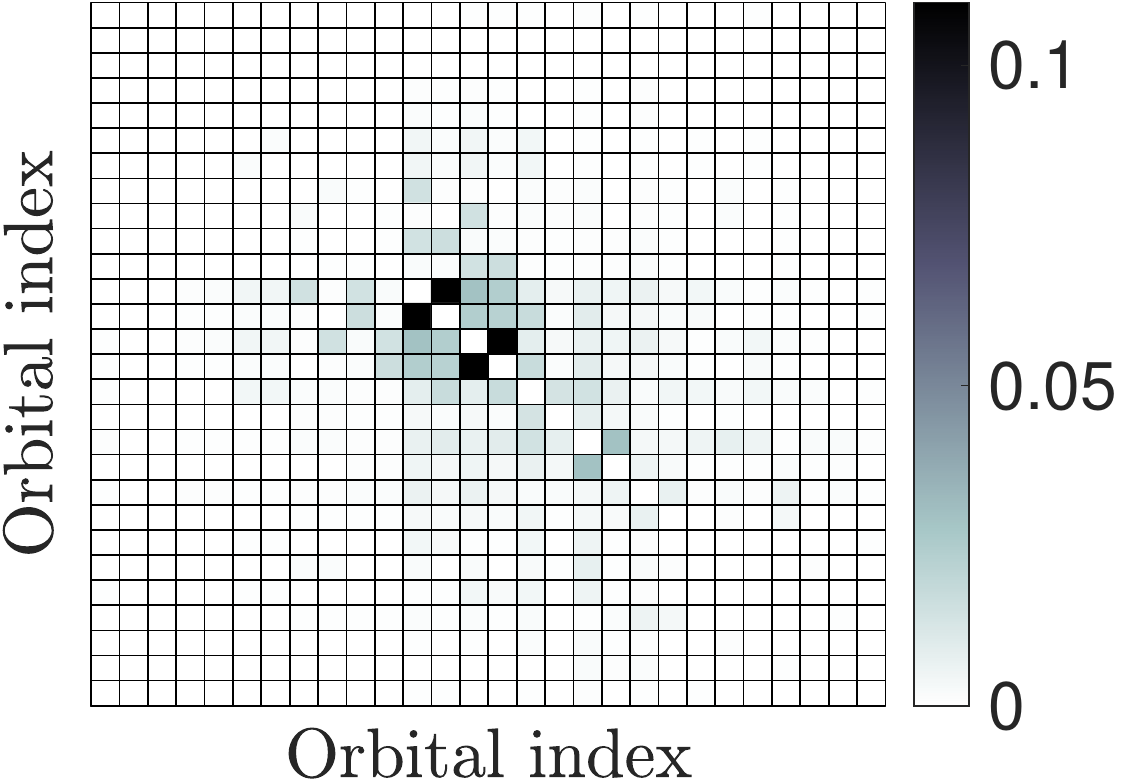}
}
\centerline{
  \includegraphics[width=0.333\textwidth]{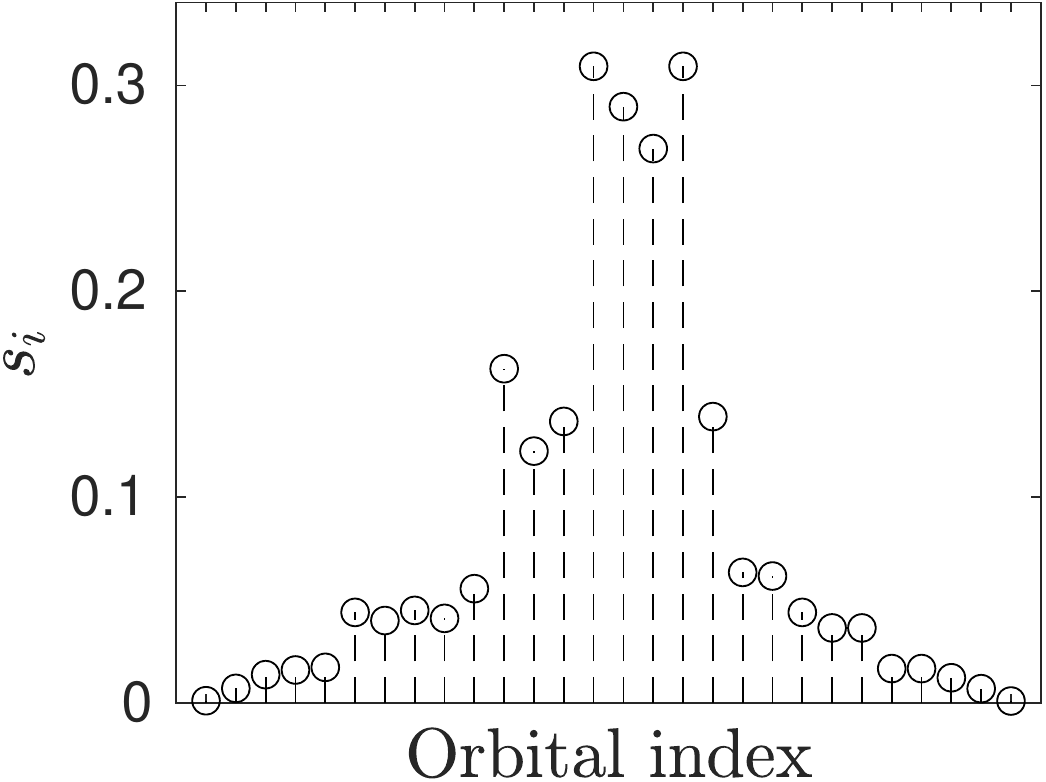}
  \includegraphics[width=0.333\textwidth]{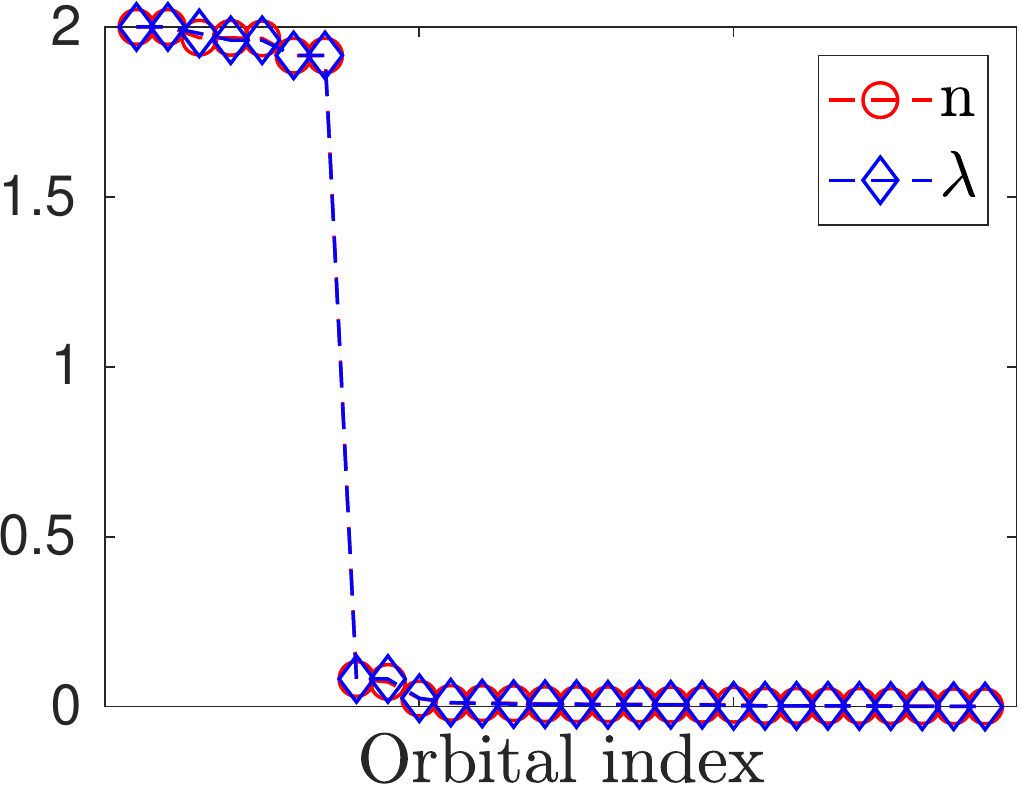}
  \includegraphics[width=0.36\textwidth]{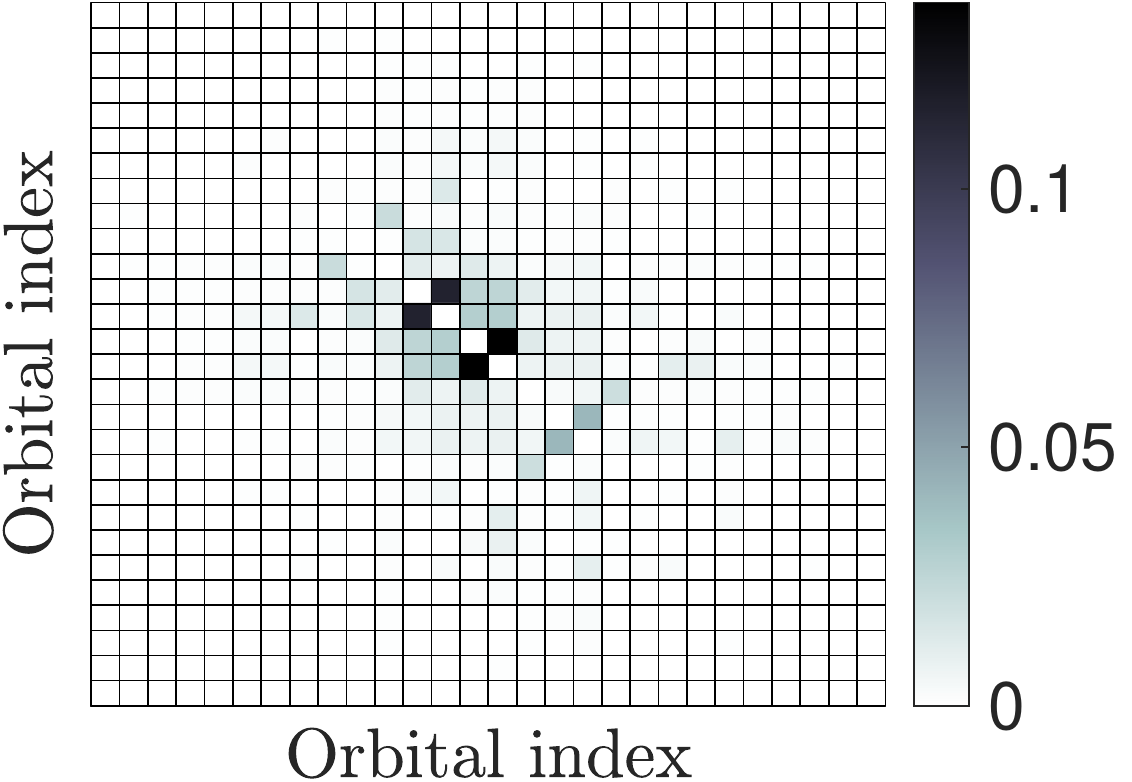}
}
\caption{Orbital entropy profiles $\{s_i\}$, sorted values of the natural orbital occupation numbers $\{\lambda_i\}$, 
occupation numbers $\{\langle n_i\rangle\}$ and mutual
informations $\{I_{i,j}\}$ for the full space CAS(14,28) (first row), and after the $20^\text{th}$ orbital optimization macro-iterations (second row) for the nitrogen dimer for bond length $r=2.118a_0$ using $D=4096$.}
\label{fig:entropies-14-28-r2118}
\end{figure}

\begin{figure}
\centerline{
  \includegraphics[width=0.333\textwidth]{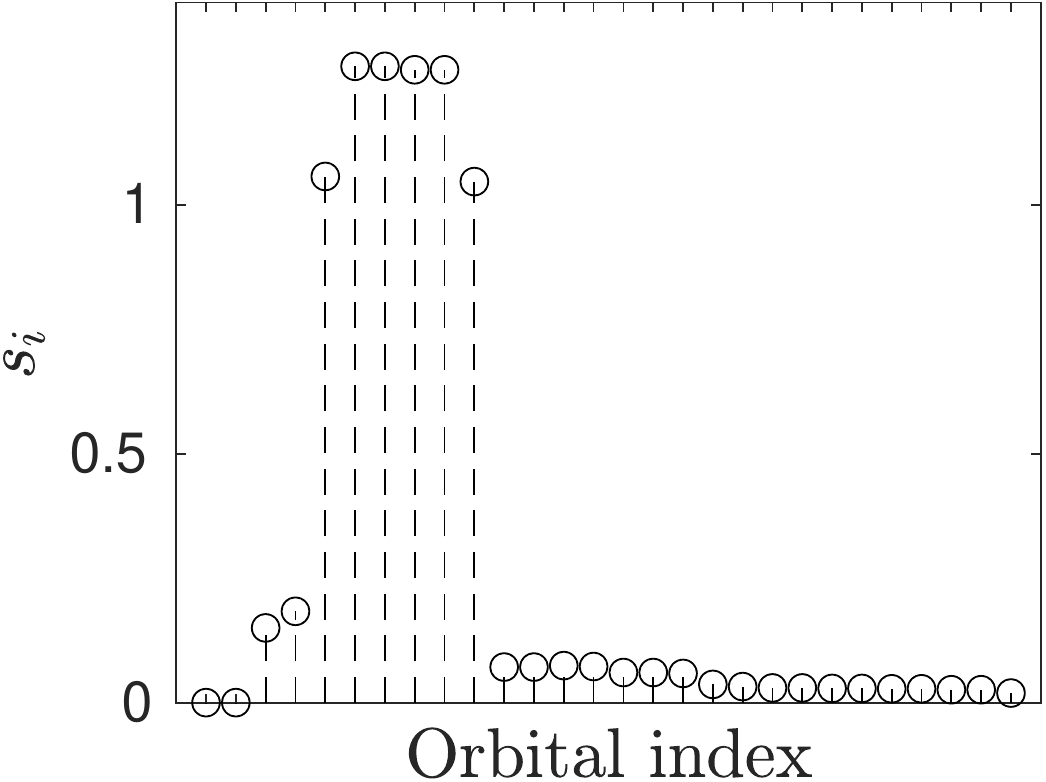}
  \includegraphics[width=0.333\textwidth]{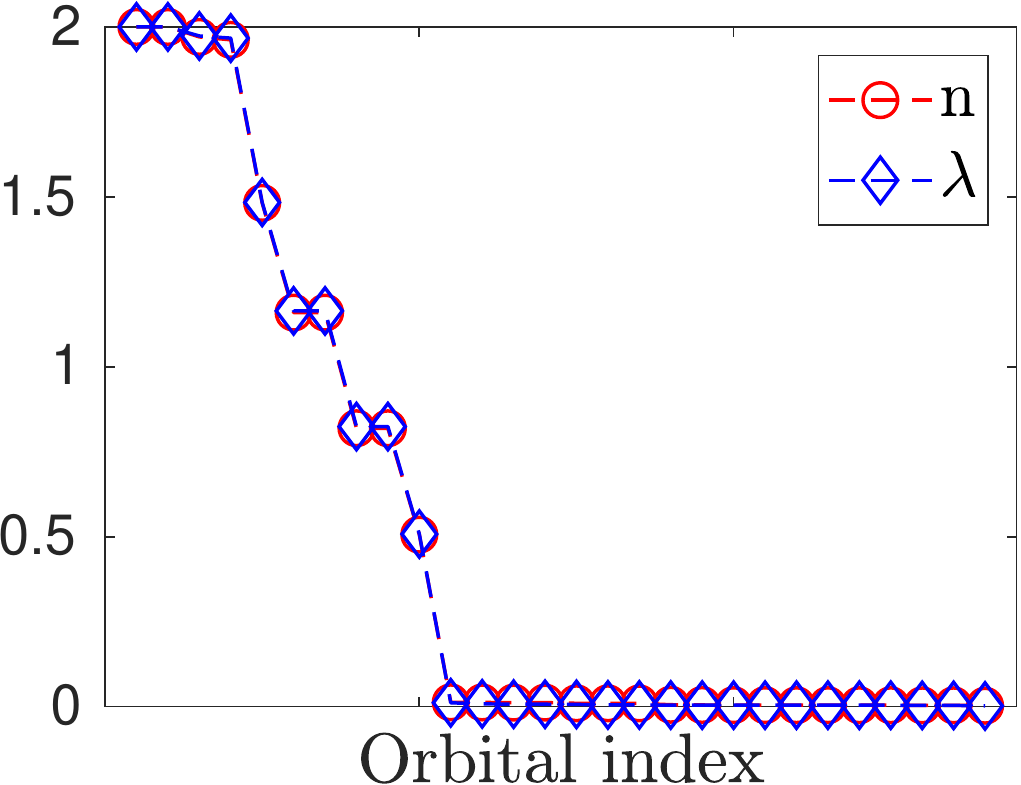}
  \includegraphics[width=0.35\textwidth]{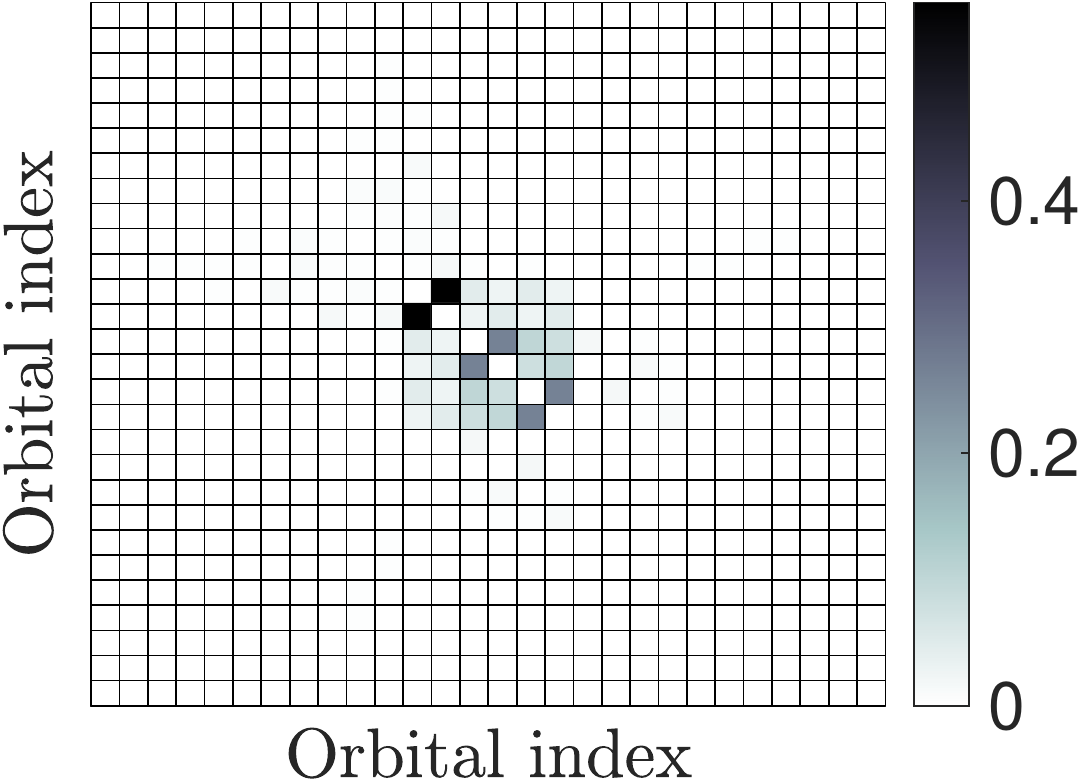}
}
\centerline{
\includegraphics[width=0.333\textwidth]{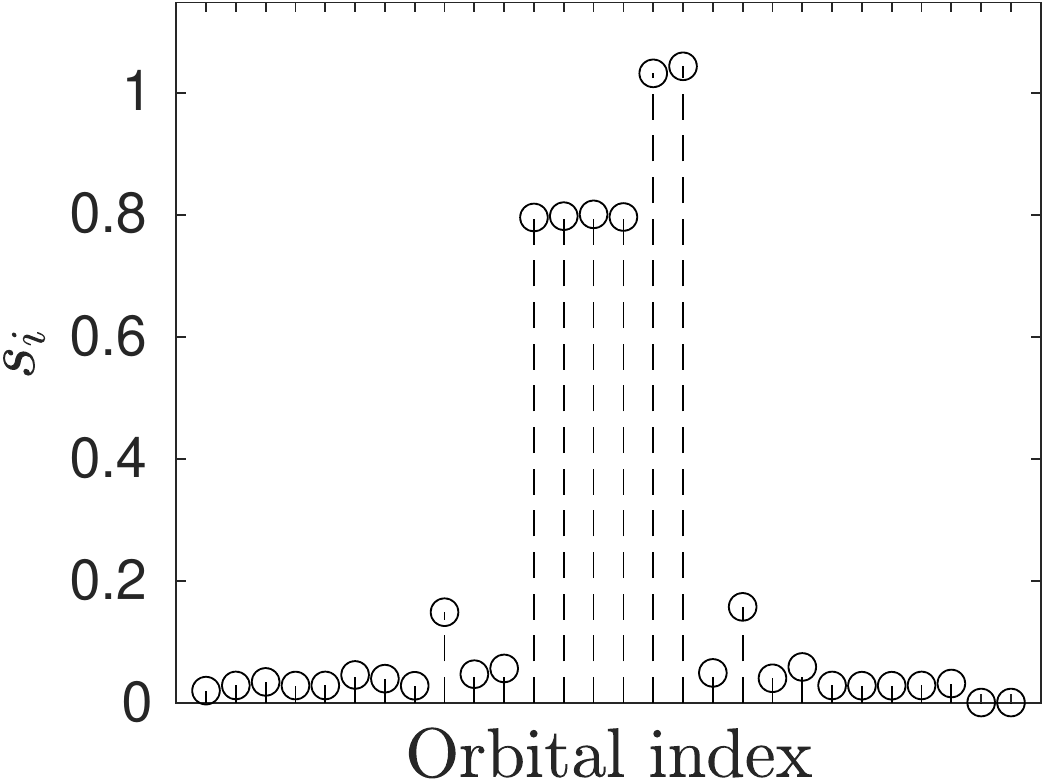}
  \includegraphics[width=0.333\textwidth]{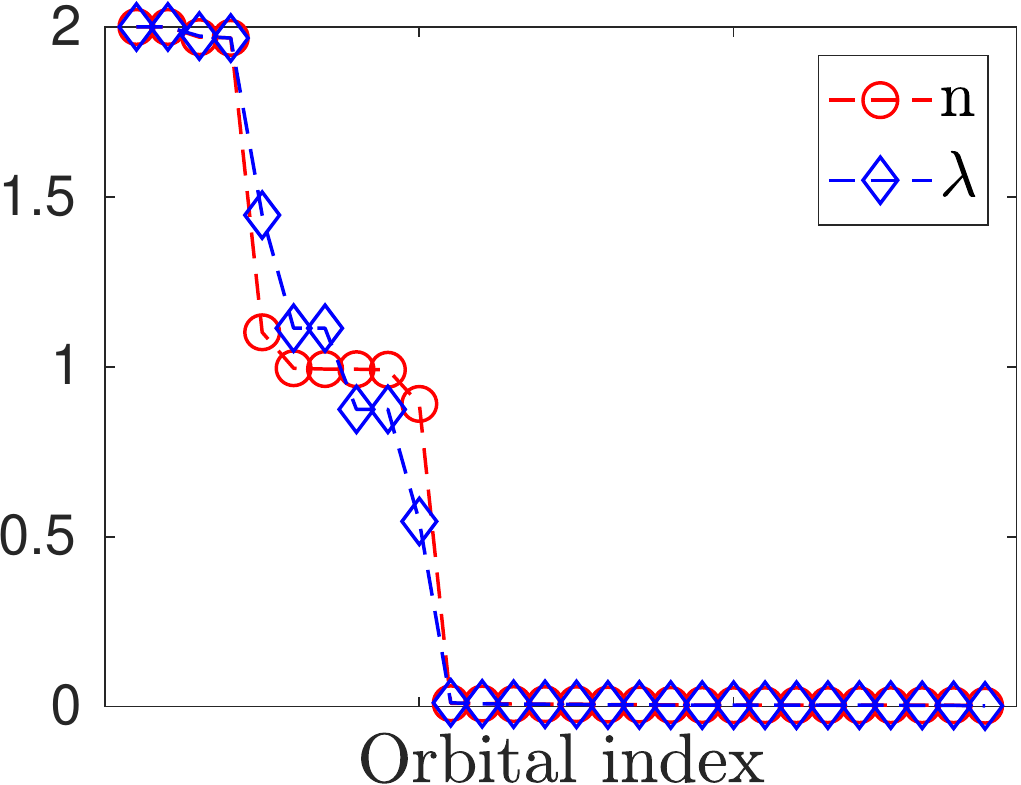}
  \includegraphics[width=0.35\textwidth]{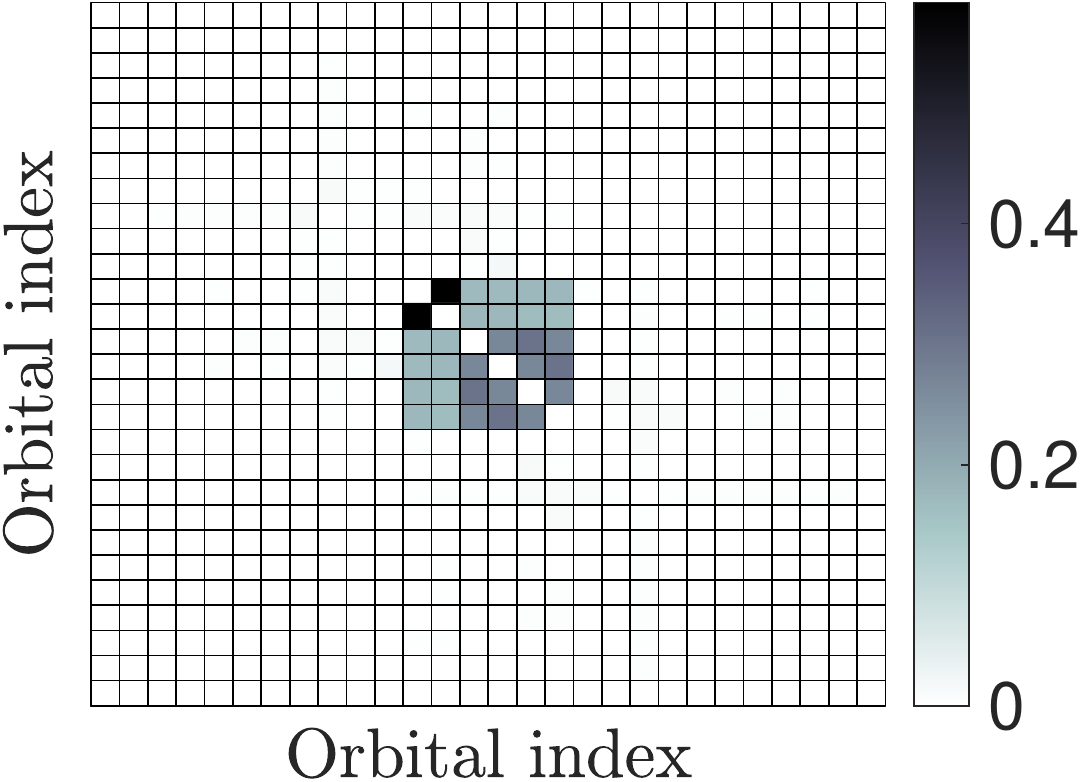}
}
\caption{(a) Similar to Fig.\ref{fig:entropies-14-28-r2118}, but for a stretched geometry at $r=4.200a_0$.}
\label{fig:entropies-14-28-r4200}
\end{figure}

In this Appendix, various selected quantities are shown 
in Figs.~\ref{fig:entropies-14-28-r2118} and~\ref{fig:entropies-14-28-r4200}
for the full space, i.e., for CAS(14,28),
to monitor the performance of the fermionic orbital optimization procedure.

\section{Additional CI based analysis}
\label{sec:appCI}

\begin{figure}
\centerline{
  \includegraphics[width = 0.49\textwidth]{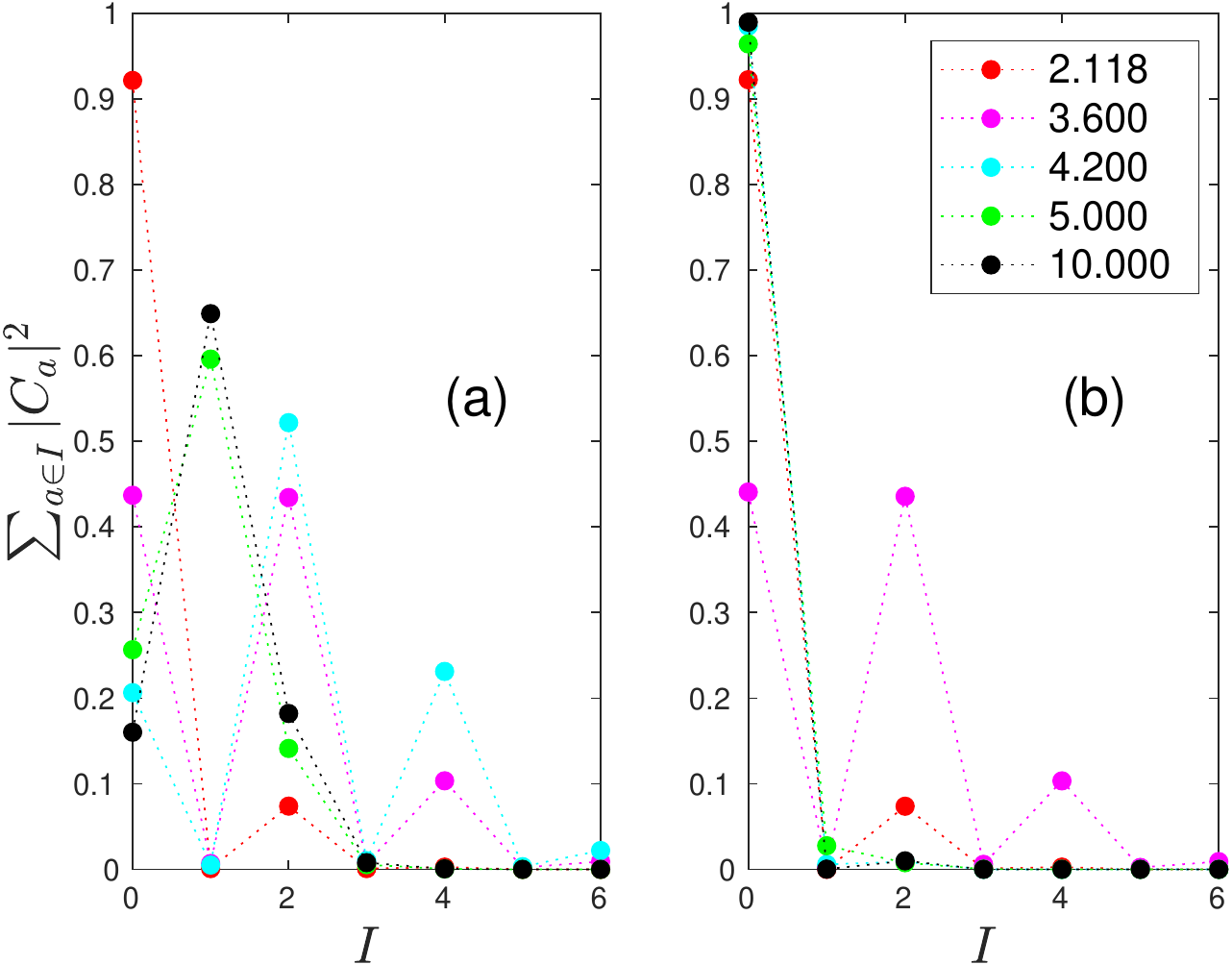}
  \includegraphics[width = 0.49\textwidth]{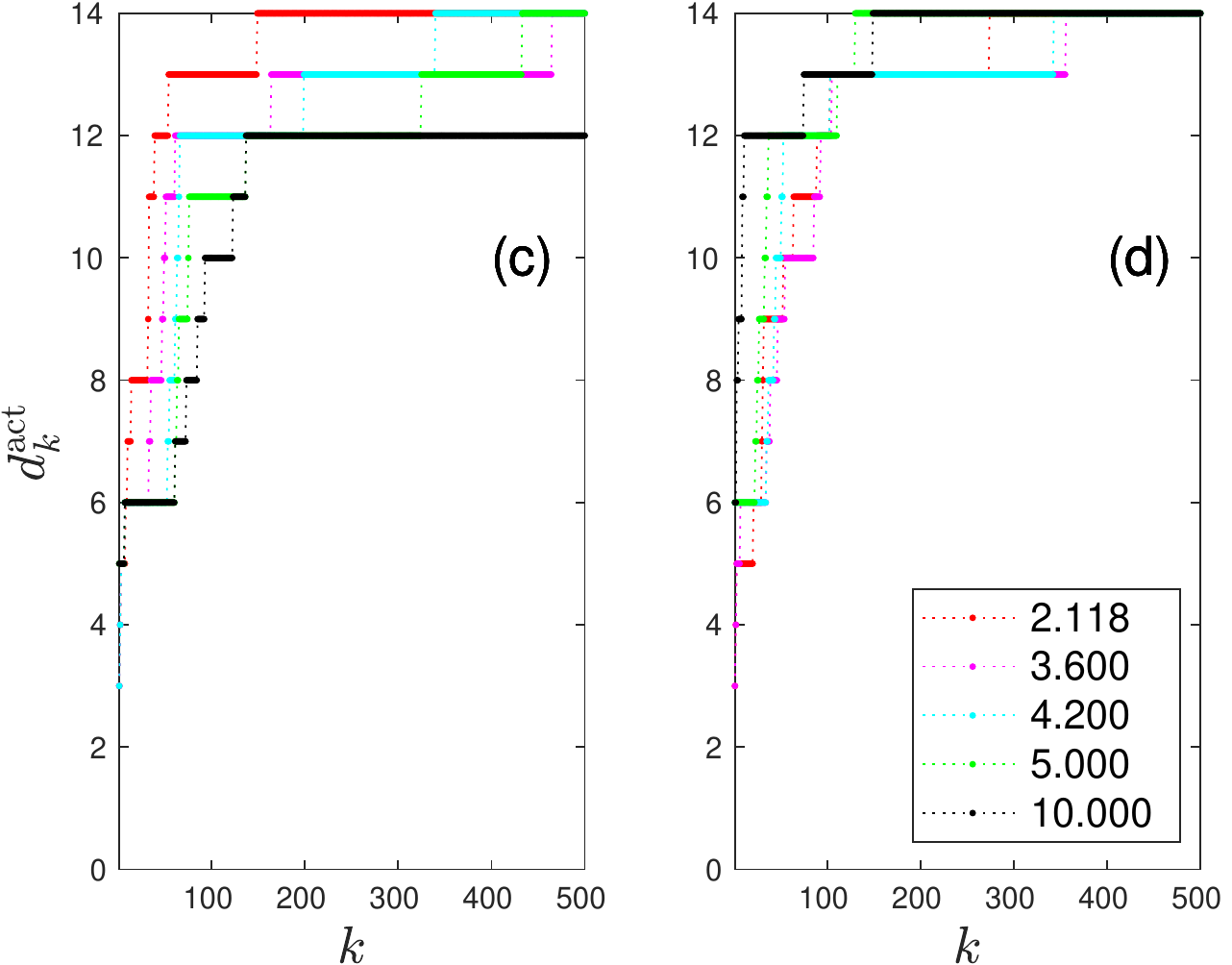}
}
\caption{(a) Square of the $C_I$ weights of the individual excitation levels
for the nitrogen dimer in the CAS(6,14) for various bond lengths, extracted from the MPS wave function, obtained by the DMRG algorithm, with a bond dimension $D=4096$. 
(b) Similar to (a), but for the optimized MOs.
(c) Number of active orbitals for the first $k$ decreasingly ordered coefficients $\vert C_a\vert$.
(d) Similar to (c), but for the optimized MOs.
The dotted lines are guides to the eye.}
\label{fig:dact}
\end{figure}

It provides further insight
to monitor the square of the $C_I$ weights of the individual excitation levels
for the nitrogen dimer in the CAS(6,14) for various bond lengths,
extracted from the MPS wave function, obtained by the DMRG algorithm, 
for the initial and optimized MOs,
shown in Fig.~\ref{fig:dact}(a) and (b) respectively.
The suppression of the $C_I$ weights for $I\neq0$ and $r\ge4.200$
is clearly visible for the optimized modes.
Fig.~\ref{fig:dact}(c) and (d) show
the number of active orbitals
for the first $k$ decreasingly ordered coefficients $\vert C_a\vert$,
for the initial and optimized MOs, respectively.
In the initial MOs for the equilibrium geometry,
the configuration $\alpha$ with highest weight $\vert C_\alpha \vert^2$ ($k=1$)
possess $6$ electrons confined on three orbitals, i.e., ($d_1^\mathrm{act}=3$),
and $d_k^\mathrm{act}$ increases to 14 for large $k$.
In contrast to this, for the stretched geometry $r=10.0$,
we have $d_k^\mathrm{act}=5$ for $k=1,2,\dots,6$, defining the active space for configuration with largest weight.
In the optimized MOs,
the picture does not change for the equilibrium geometry,
while for the two configuration with largest weight we have $d_1^\mathrm{act}=d_2^\mathrm{act}=6$. 
This information might be useful from a CI-based perspective, however,
in the MPS wave function, due to the different parametrization of the wave function, this is not so relevant.

\newpage

\bibliographystyle{unsrt}
\bibliography{paper}

\begin{thebibliography}{10}

\bibitem{White-1999}
Steven~R. White and Richard~L. Martin.
\newblock Ab initio quantum chemistry using the density matrix renormalization
  group.
\newblock {\em The Journal of Chemical Physics}, 110(9):4127--4130, 1999.

\bibitem{Legeza-2008}
{\"O}.~Legeza, R.M. Noack, J.~S\'olyom, and L.~Tincani.
\newblock Applications of quantum information in the density-matrix
  renormalization group.
\newblock In H.~Fehske, R.~Schneider, and A.~Weiße, editors, {\em
  Computational Many-Particle Physics}, volume 739 of {\em Lecture Notes in
  Physics}, pages 653--664. Springer, Berlin, Heidelberg, 2008.

\bibitem{Chan-2008}
Garnet Kin-Lic Chan, Jonathan~J. Dorando, Debashree Ghosh, Johannes Hachmann,
  Eric Neuscamman, Haitao Wang, and Takeshi Yanai.
\newblock An introduction to the density matrix renormalization group ansatz in
  quantum chemistry.
\newblock In Stephen Wilson, Peter~J. Grout, Jean Maruani, Gerardo
  Delgado-Barrio, and Piotr Piecuch, editors, {\em Frontiers in Quantum Systems
  in Chemistry and Physics}, volume~18 of {\em Progress in Theoretical
  Chemistry and Physics}. Springer, Netherlands, 2008.

\bibitem{Yanai-2009}
Takeshi Yanai, Yuki Kurashige, Debashree Ghosh, and Garnet Kin-Lic Chan.
\newblock Accelerating convergence in iterative solution for large-scale
  complete active space self-consistent-field calculations.
\newblock {\em International Journal of Quantum Chemistry}, 109(10):2178--2190,
  2009.

\bibitem{Marti-2010c}
Konrad~H. Marti and Markus Reiher.
\newblock The density matrix renormalization group algorithm in quantum
  chemistry.
\newblock {\em Zeitschrift f\"ur Physikalische Chemie}, 224:583--599, 2010.

\bibitem{Wouters-2014a}
Sebastian Wouters, Ward Poelmans, Paul~W. Ayers, and Dimitri~Van Neck.
\newblock {CheMPS2}: A free open-source spin-adapted implementation of the
  density matrix renormalization group for ab initio quantum chemistry.
\newblock {\em Computer Physics Communications}, 185(6):1501 -- 1514, 2014.

\bibitem{Legeza-2014}
{\"O}rs Legeza, Thorsten Rohwedder, Reinhold Schneider, and {\relax
  Sz}il{\'a}rd {\relax Sz}alay.
\newblock Tensor product approximation ({DMRG}) and coupled cluster method in
  quantum chemistry.
\newblock In Volker Bach and Luigi Delle~Site, editors, {\em Many-Electron
  Approaches in Physics, Chemistry and Mathematics}, Mathematical Physics
  Studies, pages 53--76. Springer International Publishing, Switzerland, 2014.

\bibitem{Szalay-2015a}
{\relax Sz}il{\'a}rd {\relax Sz}alay, Max Pfeffer, Valentin Murg, Gergely
  Barcza, Frank Verstraete, Reinhold Schneider, and {\"O}rs Legeza.
\newblock Tensor product methods and entanglement optimization for ab initio
  quantum chemistry.
\newblock {\em Int. J. Quantum Chem.}, 115(19):1342--1391, 2015.

\bibitem{Chan-2016}
Garnet Kin-Lic Chan, Anna Keselman, Naoki Nakatani, Zhendong Li, and Steven~R.
  White.
\newblock Matrix product operators, matrix product states, and ab initio
  density matrix renormalization group algorithms.
\newblock {\em The Journal of Chemical Physics}, 145(1):014102, 2016.

\bibitem{Baiardi-2020}
Alberto Baiardi and Markus Reiher.
\newblock The density matrix renormalization group in chemistry and molecular
  physics: Recent developments and new challenges.
\newblock {\em The Journal of Chemical Physics}, 152(4):040903, 2020.

\bibitem{Cheng-2022}
Yifan Cheng, Zhaoxuan Xie, and Haibo Ma.
\newblock Post-density matrix renormalization group methods for describing
  dynamic electron correlation with large active spaces.
\newblock {\em The Journal of Physical Chemistry Letters}, 13(3):904--915,
  2022.

\bibitem{Legeza-2003a}
{\"O}.~Legeza, J.~R\"oder, and B.~A. Hess.
\newblock Controlling the accuracy of the density-matrix renormalization-group
  method: The dynamical block state selection approach.
\newblock {\em Phys. Rev. B}, 67:125114, Mar 2003.

\bibitem{Holtz-2012a}
Sebastian Holtz, Thorsten Rohwedder, and Reinhold Schneider.
\newblock On manifolds of tensors of fixed {TT}-rank.
\newblock {\em Numerische Mathematik}, 120(4):701--731, 2012.

\bibitem{Holtz-2012b}
S.~Holtz, T.~Rohwedder, and R.~Schneider.
\newblock The alternating linear scheme for tensor optimization in the tensor
  train format.
\newblock {\em SIAM Journal on Scientific Computing}, 34(2):A683--A713, 2012.

\bibitem{Legeza-2003b}
{\"O}.~Legeza and J.~S\'olyom.
\newblock Optimizing the density-matrix renormalization group method using
  quantum information entropy.
\newblock {\em Phys. Rev. B}, 68:195116, Nov 2003.

\bibitem{Nakatani-2013}
Naoki Nakatani and Garnet Kin-Lic Chan.
\newblock Efficient tree tensor network states ({TTNS}) for quantum chemistry:
  Generalizations of the density matrix renormalization group algorithm.
\newblock {\em The Journal of Chemical Physics}, 138(13):134113, 2013.

\bibitem{Murg-2015}
Valentin Murg, Frank Verstraete, Reinhold Schneider, P{\'e}ter~R. Nagy, and
  {\"O}rs Legeza.
\newblock Tree tensor network state with variable tensor order: An efficient
  multireference method for strongly correlated systems.
\newblock {\em Journal of Chemical Theory and Computation}, 11(3):1027--1036,
  2015.

\bibitem{Gunst-2018}
Klaas Gunst, Frank Verstraete, Sebastian Wouters, {\"O}rs Legeza, and Dimitri
  Van~Neck.
\newblock {T3NS}: Three-legged tree tensor network states.
\newblock {\em Journal of Chemical Theory and Computation}, 14(4):2026--2033,
  2018.

\bibitem{Rissler-2006}
J\"org Rissler, Reinhard~M. Noack, and Steven~R. White.
\newblock Measuring orbital interaction using quantum information theory.
\newblock {\em Chemical Physics}, 323(2–3):519 -- 531, 2006.

\bibitem{Murg-2010a}
V.~Murg, F.~Verstraete, {\"O}.~Legeza, and R.~M. Noack.
\newblock Simulating strongly correlated quantum systems with tree tensor
  networks.
\newblock {\em Phys. Rev. B}, 82:205105, Nov 2010.

\bibitem{Stein-2016}
Christopher~J. Stein and Markus Reiher.
\newblock Automated selection of active orbital spaces.
\newblock {\em Journal of Chemical Theory and Computation}, 12(4):1760--1771,
  2016.

\bibitem{Fertitta-2014}
E.~Fertitta, B.~Paulus, G.~Barcza, and {\"O}.~Legeza.
\newblock Investigation of metal-insulator-like transition through the ab
  initio density matrix renormalization group approach.
\newblock {\em Phys. Rev. B}, 90:245129, Dec 2014.

\bibitem{Krumnow-2016}
C.~Krumnow, L.~Veis, {\"O}.~Legeza, and J.~Eisert.
\newblock Fermionic orbital optimization in tensor network states.
\newblock {\em Phys. Rev. Lett.}, 117:210402, Nov 2016.

\bibitem{Krumnow-2021}
C.~Krumnow, L.~Veis, J.~Eisert, and {\"O}.~Legeza.
\newblock Effective dimension reduction with mode transformations: Simulating
  two-dimensional fermionic condensed matter systems with matrix-product
  states.
\newblock {\em Phys. Rev. B}, 104:075137, Aug 2021.

\bibitem{Foster-1960}
J.~M. Foster and S.~F. Boys.
\newblock Canonical configurational interaction procedure.
\newblock {\em Rev. Mod. Phys.}, 32:300--302, Apr 1960.

\bibitem{Boys-1960}
S.~F. Boys.
\newblock Construction of some molecular orbitals to be approximately invariant
  for changes from one molecule to another.
\newblock {\em Rev. Mod. Phys.}, 32:296--299, Apr 1960.

\bibitem{Pipek-1989}
J\'anos Pipek and Paul~G. Mezey.
\newblock A fast intrinsic localization procedure applicable for ab initio and
  semiempirical linear combination of atomic orbital wave functions.
\newblock {\em The Journal of Chemical Physics}, 90(9):4916--4926, 1989.

\bibitem{Dunning-1989}
Thom~H. Dunning.
\newblock Gaussian basis sets for use in correlated molecular calculations.
  {I}. {T}he atoms boron through neon and hydrogen.
\newblock {\em The Journal of Chemical Physics}, 90(2):1007--1023, 1989.

\bibitem{Helgaker-2000}
Trygve Helgaker, Poul Jorgensen, and Jeppe Olsen.
\newblock {\em Molecular electronic-structure theory}.
\newblock Wiley New York, 2000.

\bibitem{Vidal-2003b}
Guifr\'e Vidal.
\newblock Efficient classical simulation of slightly entangled quantum
  computations.
\newblock {\em Phys. Rev. Lett.}, 91:147902, Oct 2003.

\bibitem{Ostlund-1995}
Stellan \"Ostlund and Stefan Rommer.
\newblock Thermodynamic limit of density matrix renormalization.
\newblock {\em Phys. Rev. Lett.}, 75:3537--3540, Nov 1995.

\bibitem{Verstraete-2004a}
F.~Verstraete and J.~I. Cirac.
\newblock Renormalization algorithms for quantum-many body systems in two and
  higher dimensions.
\newblock {\em arXiv [cond-mat.str-el]}, page 0407066, 2004.

\bibitem{Verstraete-2004b}
F.~Verstraete, D.~Porras, and J.~I. Cirac.
\newblock Density matrix renormalization group and periodic boundary
  conditions: A quantum information perspective.
\newblock {\em Phys. Rev. Lett.}, 93:227205, Nov 2004.

\bibitem{Szalay-2021}
{\relax Sz}il{\'a}rd {\relax Sz}alay, Zolt{\'a}n Zimbor{\'a}s, Mih{\'a}ly
  M{\'a}t{\'e}, Gergely Barcza, Christian Schilling, and {\"O}rs Legeza.
\newblock Fermionic systems for quantum information people.
\newblock {\em Journal of Physics A: Mathematical and Theoretical},
  54(39):393001, 2021.

\bibitem{Boguslawski-2013}
Katharina Boguslawski, Pawe\l Tecmer, Gergely Barcza, {\"O}rs Legeza, and
  Markus Reiher.
\newblock Orbital entanglement in bond-formation processes.
\newblock {\em Journal of Chemical Theory and Computation}, 9(7):2959--2973,
  2013.

\bibitem{Szalay-2015b}
{\relax Sz}il\'ard {\relax Sz}alay.
\newblock Multipartite entanglement measures.
\newblock {\em Phys. Rev. A}, 92:042329, Oct 2015.

\bibitem{Szalay-2019}
{\relax Sz}il{\'a}rd {\relax Sz}alay.
\newblock $k$-stretchability of entanglement, and the duality of
  $k$-separability and $k$-producibility.
\newblock {\em Quantum}, 3:204, 2019.

\bibitem{Szalay-2017}
{\relax Sz}il{\'a}rd {\relax Sz}alay, Gergely Barcza, Tibor {\relax
  Sz}ilv{\'a}si, Libor Veis, and {\"O}rs Legeza.
\newblock The correlation theory of the chemical bond.
\newblock {\em Scientific Reports}, 7:2237, May 2017.

\bibitem{Brandejs-2019}
Jan Brandejs, Libor Veis, {\relax Sz}il{\'a}rd {\relax Sz}alay, Ji{\u r}{\'i}
  Pittner, and {\"O}rs Legeza.
\newblock Quantum information-based analysis of electron-deficient bonds.
\newblock {\em The Journal of Chemical Physics}, 150(20):204117, 2019.

\bibitem{Barcza-2011}
G.~Barcza, {\"O}.~Legeza, K.~H. Marti, and M.~Reiher.
\newblock Quantum-information analysis of electronic states of different
  molecular structures.
\newblock {\em Phys. Rev. A}, 83:012508, Jan 2011.

\end{thebibliography}

\end{document}